\documentclass[%
 reprint, 
superscriptaddress,
 amsmath,amssymb,
 aps,physrev,
floatfix,
]{revtex4-1}

\usepackage{times}
\usepackage[pdftex]{graphicx}
\usepackage[bookmarksopen=true,bookmarksopenlevel=1]{hyperref}
\usepackage{epstopdf}
\usepackage{amsmath}
\usepackage{amsfonts}
\usepackage{amssymb}
\usepackage{graphicx}
\usepackage{sidecap}
\sidecaptionvpos{figure}{c}
\usepackage{color}
\usepackage{dcolumn}
\usepackage{bm}
\usepackage[compact]{titlesec}
\usepackage[normalem]{ulem}
\usepackage{bookmark}
\usepackage{natbib}
\usepackage{indentfirst}
\usepackage{appendix}
\usepackage{soul}

\usepackage{lineno}

\definecolor{mypink}{RGB}{219, 48, 122}
\definecolor{mygreen}{RGB}{51, 153, 102}
\definecolor{brown}{RGB}{165, 42, 42}

\newlength{\beforesection}
\newlength{\aftersection}
\newlength{\beforesubsection}
\newlength{\aftersubsection}
\titlespacing*{\section}{0pt}{\beforesection}{\aftersection}
\titlespacing*{\subsection}{0pt}{\beforesubsection}{\aftersubsection}

\def\invA{\AA$^{-1}$}

\def\invA{\AA$^{-1}$}

\def\root33{$\sqrt{3}\times\sqrt{3}$ {\it R}30$^\circ$}
\def\RT3{$\sqrt{3}$}

\begin{document}

\title{Spin-polarised surface fermiology of ohmic WSe$_2$/NbSe$_2$ interfaces}

\author{Oliver J. Clark}
\email[Corresponding author. E-mail address: ] {oliver.clark@diamond.ac.uk}
\affiliation{School of Physics and Astronomy, Monash University, Clayton, VIC, Australia}

\author{Thi-Hai-Yen Vu}
\affiliation{School of Physics and Astronomy, Monash University, Clayton, VIC, Australia}

\author{Ben A. Chambers}
\affiliation{Flinders Microscopy and Microanalysis, Flinders University, Bedford Park, South Australia 5042, Australia}

\author{Federico Mazzola}
\affiliation{Department of Molecular Sciences and Nanosystems,
Ca Foscari University of Venice, Venice IT-30172, Italy}
\affiliation{CNR-SPIN UOS Napoli, Complesso Universitario di
Monte Sant’Angelo, Via Cinthia 80126, Napoli, Italy}

\author{Sadhana Sridhar}
\affiliation{School of Physics and Astronomy, Monash University, Clayton, VIC, Australia}

\author{Geetha Balakrishnan}
\affiliation{Department of Physics, University of Warwick, Coventry CV4 7AL, United Kingdom}

\author{Aaron Bostwick}
\affiliation{Advanced Light Source, Lawrence Berkeley National Laboratory, Berkeley, CA, 94720 USA}

\author{Chris Jozwiak}
\affiliation{Advanced Light Source, Lawrence Berkeley National Laboratory, Berkeley, CA, 94720 USA}

\author{Eli Rotenberg}
\affiliation{Advanced Light Source, Lawrence Berkeley National Laboratory, Berkeley, CA, 94720 USA}

\author{Sarah L. Harmer}
\affiliation{Institute for Nanoscale Science and Technology, Flinders University, Bedford Park, South Australia 5042, Australia}
\affiliation{Flinders Microscopy and Microanalysis, Flinders University, Bedford Park, South Australia 5042, Australia}

\author{Michael S. Fuhrer}
\affiliation{School of Physics and Astronomy, Monash University, Clayton, VIC, Australia}
\affiliation{ARC Centre for Future Low Energy Electronics Technologies, Monash University, Clayton, VIC, Australia}

\author{Mark T. Edmonds}
\affiliation{School of Physics and Astronomy, Monash University, Clayton, VIC, Australia}
\affiliation{ARC Centre for Future Low Energy Electronics Technologies, Monash University, Clayton, VIC, Australia}
\affiliation{ANFF-VIC Technology Fellow, Melbourne Centre for Nanofabrication, Victorian Node of the Australian National Fabrication Facility, Clayton, VIC 3168, Australia}


\begin{abstract}
Discovering and engineering spin-polarised surface states in the electronic structures of condensed matter systems is a crucial first step in development of spintronic devices, wherein spin-polarised bands crossing the Fermi level can facilitate information transfer. Here, we show how the spin-orbit split K-point valleys of monolayer WSe$_2$ can be made potentially suitable for this purpose, despite the semiconducting ground state. By interfacing with metallic 2H-NbSe$_2$, these valence band extrema are shifted by $\sim$800~meV to produce a surface-localised Fermi surface populated only by spin-polarised carriers. By increasing the WSe$_2$ thickness, the Fermi pockets can be moved from K to $\Gamma$, demonstrating tunability of novel semi-metallic phases that exist atop a substrate additionally possessing charge density wave and superconducting transitions. Together, this study provides spectroscopic understanding into $p$-type, Schottky barrier-free interfaces, which are of urgent interest for bypassing the limitations of  current-generation vertical field effect transistors, in addition to longer-term spintronics development.
\end{abstract}

\maketitle

\section{\label{sec:intro}INTRODUCTION\protect\\ }

As the limitations of conventional electronics are reached, the semiconducting subset of 2H-structured transition metal dichalcogenides (TMDs) have emerged as key components for ultra-optimized nanoscale electronics due to the ease in which their monolayer forms can be isolated without `dangling' bonds or residual polar charge. This facilitates, for example, the fabrication of vertical field effect transistors (VFETs) on the smallest possible length scales through interfacing with  metallic 2D materials, like graphene~\cite{patoary_improvements_2023,park_p_2025,sata_n_2017,hu_interface_2025}. 

In parallel, spintronics are in active development, relying on the spin of electrons rather than charge. Spintronic devices are underpinned by by materials in which the states near the Fermi level are spin polarized and responsive to external stimuli~\cite{datta_electronic_1990, gorkov_superconducting_2001,lesne_highly_2016, manchon_new_2015,hirohata_review_2020, breunig_opportunities_2022}. In non-magnetic materials, Rashba-type spin-orbit coupling is a key mechanism for the lifting of the spin degeneracy of a band, producing a spin polarization orthogonal to the direction of an electric dipole in real space~\cite{Rashba84}.  The larger the separation of up and down spin species in energy-momentum space, the  more resilient the system is to back scattering, and therefore the more useful practically~\cite{ogowa_photocontrol_2014, sanchez_spin_2013, manchon_new_2015, jungfleisch_control_2018, varotto_room_2021,rinaldi_ferroelectric_2018}. 

Semiconducting TMDs have also gained traction for this purpose: The band structure of 2H-WSe$_2$, like other 2H-TMDs, naturally possesses pairs of spin-orbit split hole-like bands at the K and K' points of its Brillouin zone (BZ). In the monolayer limit, shown in Fig.~\ref{Fig1}(a), these pairs of bands are spin-polarised out-of-plane due to an uncompensated in-plane electric dipole within a single 1H-unit~\cite{xaio_coupled_2012, zeng_valley_2012, mak_control_2012, riley_direct_2014}, but reverse from K to K' with time-reversal symmetry (TRS). Together, this produces oppositely-polarized valence band maxima (VBM) with maximal separation in momentum space~\cite{xaio_coupled_2012, zeng_valley_2012, mak_control_2012, riley_direct_2014}. While this band structure has already been exploited in so-called valleytronic-devices, wherein electrons populating these Zeeman-like VBM can be selectively excited with circularly polarised light~\cite{xaio_coupled_2012, zeng_valley_2012, mak_control_2012, yuan_zeeman_2013}, the energetic positioning of these bands is $\sim$-0.75~eV below the Fermi level, restricting spintronic functionality.

Here, through nano-focussed angle-resolved photoemisison (nano-ARPES), we show how these spin-split valence bands can be moved to the Fermi level by interfacing with a metallic 2H-NbSe$_2$ crystal. These ohmic interfaces (i.e. with negative Schotkky barriers) thereby produce an artificial surface electronic structure composed only of spin-polarised bands at the Fermi level. 
This study therefore provides a microscopic, electronic structure perspective of $p-$type NbSe$_2$/WSe$_2$ VFETS, while indicating potential ways to harness the unique electronic spin structures of 2H-TMDs at equilibrium and without optical pumping.

\section{\label{sec:results}RESULTS\protect\\ }

\begin{figure}
	\centering
	\includegraphics[width=\columnwidth]{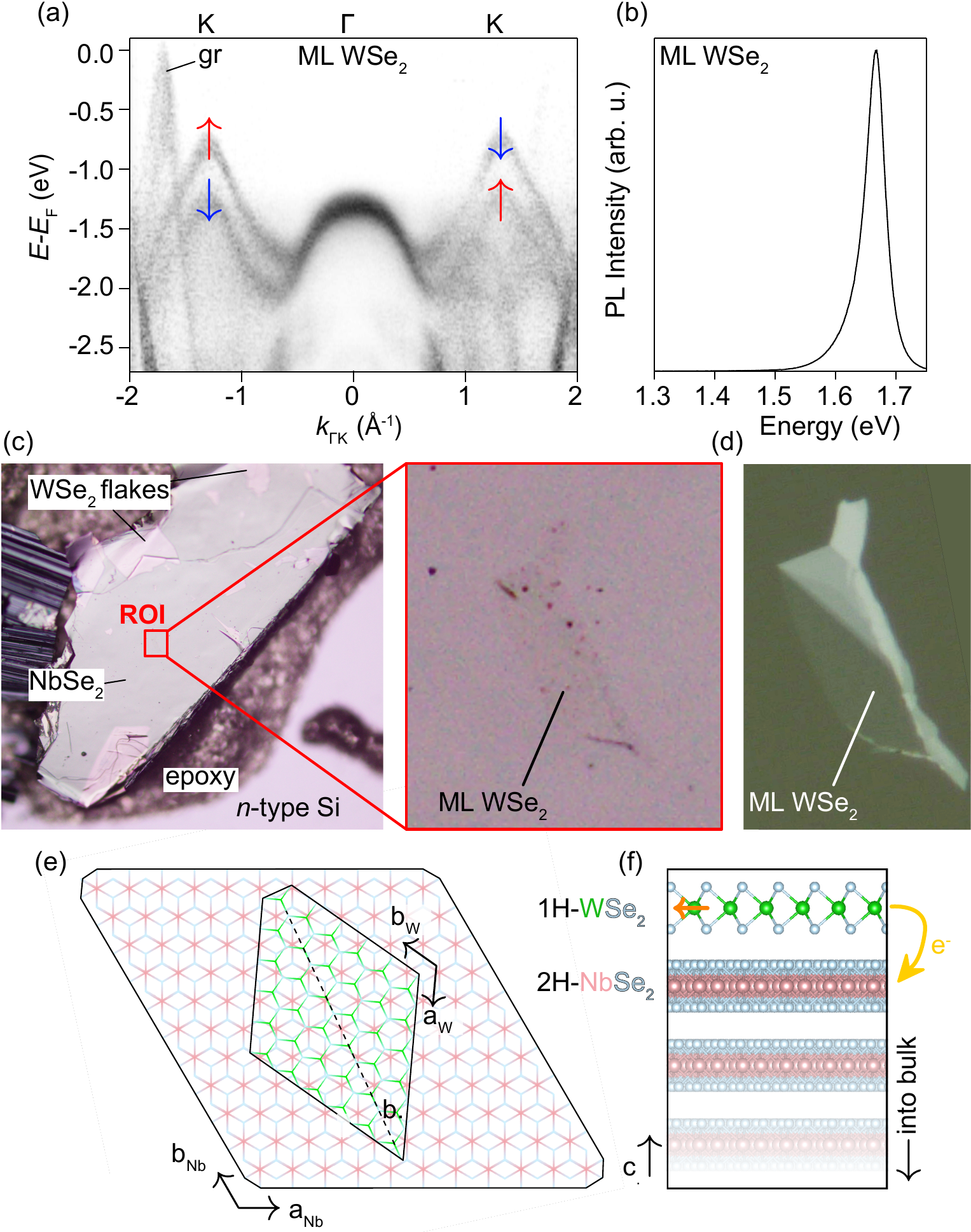}
	\caption{\label{Fig1} \text{Sample fabrication and electronic structure of monolayer WSe$_2$} (a) K'-$\Gamma$-K band dispersion (h$\nu$=21.2, T=300~K) of a second monolayer of WSe$_2$ deposited on graphite/Si substrates acquired with nano-ESCA. The out-of-plane spin polarization of the pair of SOC-split bands at $\Gamma$ are shown, as determined in earlier works~\cite{riley_direct_2014}. The band labeled gr originates from the Dirac cone of the weakly-interacting graphene/Si substrate. (b) Photoluminescence (PL) signal from the monolayer region of the flake in (d). (c) Optical image of a WSe$_2$/(NbSe$_2$)$_n$ sample. Red box indicates the area where the WSe$_2$ monolayer was transferred, shown right. (d) Flake containing the ML-WSe$_2$ region before transferring onto bulk NbSe$_2$. (e-f) Schematic diagram of a monolayer of 2H-WSe$_2$ placed upon bulk 2H-NbSe$_2$ with large finite twist angle. The dashed line in the a-b plane in (a) indicates the cross section shown in (b). Orange arrow in the WSe$_2$ layer indicates the electric dipole giving rise to the Zeeman-like out-of-plane spin polarisation indicated in (a). The yellow curved arrow shows the direction of charge transfer in this hybrid system.}
\end{figure}

\begin{figure*}[t]
	\centering	\includegraphics[width=\textwidth]{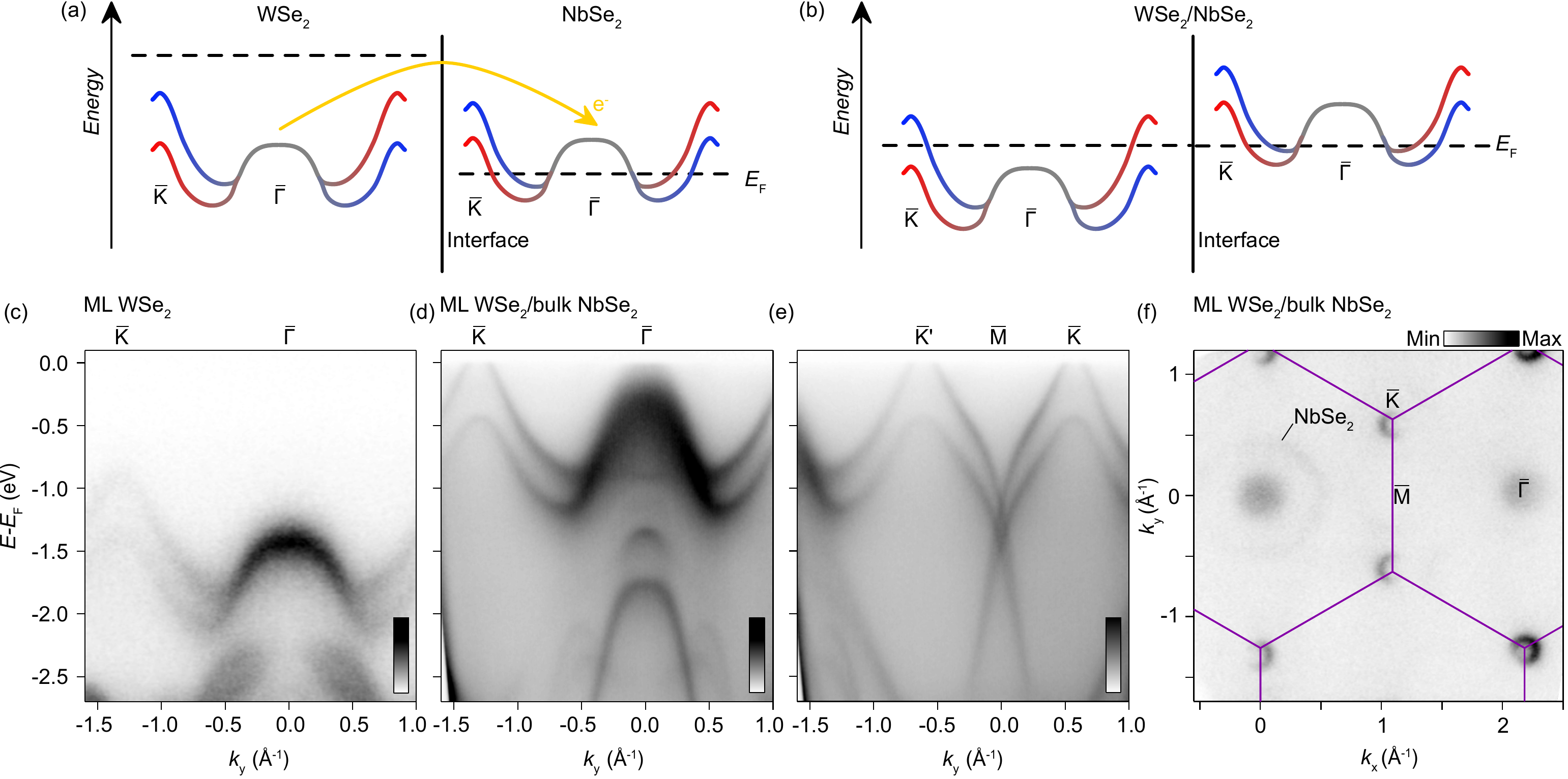}
	\caption{\label{Fig2} \text{Fermiology of ML-WSe$_2$ on bulk NbSe$_2$} (a-b) Schematic diagrams of freestanding (a) and interfaced (b) WSe$_2$ and NbSe$_2$ systems. The band colouring indicates the out-of-plane spin polarisation of the bands, though we note that for 2+ layer systems, the spin-polarisation is quenched by the adjacent layer in the unit cell. (b) Band dispersion ($h\nu=$114~eV) along the $\overline{\text{K}}$-$\overline{\Gamma}$-$\overline{\text{K}}$ for a free standing WSe$_2$, measured under similar experimental conditions to that in (d). (d-e) Band dispersions ($h\nu=$91.5~eV) along the $\overline{\text{K}}$-$\overline{\Gamma}$-$\overline{\text{K}}$ (d) and $\overline{\text{K}}$-$\overline{\text{M}}$-$\overline{\text{K}}$  (e) directions of ML-WSe$_2$ placed on a bulk NbSe$_2$ substrate. (f) Corresponding Fermi surface ($h\nu=$91.5~eV). Solid lines show the BZ of WSe$_2$. }
\end{figure*}

\begin{figure*}[t]
	\centering	\includegraphics[width=\textwidth]{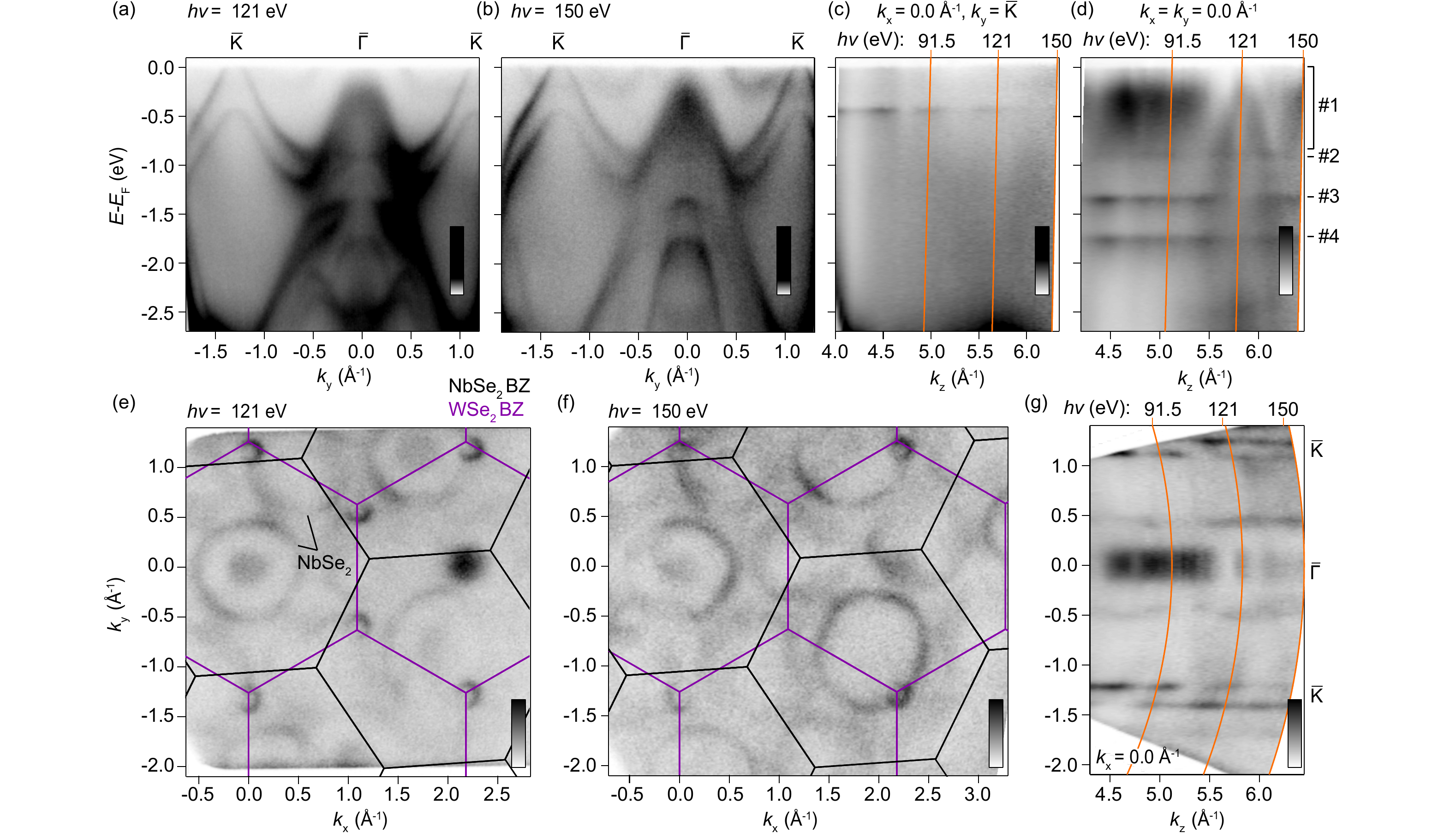}
	\caption{\label{Fig3} \text{Mixed band dimensionality in ML-WSe$_2$/bulk NbSe$_2$} (a-b) Band dispersions  along the $\overline{\text{K}}$-$\overline{\Gamma}$-$\overline{\text{K}}$ directions for $h\nu=$121~eV (a) and (b) 150~eV. (c-d) $k_z$-dispersions through the $\overline{\text{K}}$ point (c) and $\overline{\Gamma}$ point (d) constructed by varying the incident photon energy between 60 and 150~eV ($V_0$=13~eV). (e-f) Fermi surface maps taken with $h\nu=$121~eV (e) and 150~eV (f). Black and purple solid lines are Brilluoin zone contours for WSe$_2$ and NbSe$_2$, respectively. (g) $k_y$-$k_z$ constant energy contour at the Fermi level constructed by varying the incident photon energy between 60 and 150~eV ($V_0$=13~eV). Orange lines in (c,d,g) indicate the $k_z$ positioning of the 91.5~eV, 121~eV and 150~eV datasets shown in Figs.~\ref{Fig2} and~\ref{Fig3}. The four distinct bands in (d) are labeled.
    } 
\end{figure*}

Figure~\ref{Fig1}(b-f) overviews the construction and composition of a hybrid system made by pairing a monolayer WSe$_2$ flake to a bulk NbSe$_2$ crystal. Flakes of WSe$_2$ containing monolayer regions (Fig.~\ref{Fig1}(d))  were exfoliated and verified through their photoluminescence spectra (Fig.~\ref{Fig1}(b)) which is strongly thickness-dependent for semiconducting 2H-TMDs~\cite{tonndorf_photoluminesence_2013}. The flake is then transferred onto cleaved surfaces of NbSe$_2$ single crystals (Fig.~\ref{Fig1}(c)) in an Ar glovebox. A large twist angle is present in our samples (schematized in Fig.~\ref{Fig1}(e-f)) to decouple Moir\'e effects. Despite the majority of the NbSe$_2$ surface remaining exposed, the monolayer WSe$_2$ cap is found sufficient to locally preserve the underlying substrate from oxidation for several days in atmospheric conditions. 

The crystal and electronic structures of 2H-WSe$_2$ and 2H-NbSe$_2$ are qualitatively very similar. However, the Nb $d^1$ configuration (compared to W $d^2$) positions the NbSe$_2$ Fermi level below both $\Gamma$ and K local valence band maxima (lVBM), creating a metallic Fermi surface with large electron and hole sheets~\cite{chhowalla_chemistry_2013, borisenko_two_2009, bawden_spin_2015}. As schematized in Fig.~\ref{Fig2}(a-b), the monolayer WSe$_2$ top layer will lose electrons to the bulk NbSe$_2$ when interfaced, thus becoming $p$-doped. Recent transport studies have demonstrated that the valence band maximum of few-layer ($>1$) WSe$_2$ and the work function of NbSe$_2$ align in such a way to permit near-perfect $p$-type ohmic contacts between these materials~\cite{park_p_2025,sata_n_2017}. Due to the van der Waals nature of both components, there are no interface states to pin the Fermi level to the gap of the semiconductor~\cite{park_p_2025,sata_n_2017}. However, these transport studies could not rule out a small positive Schottky barrier at the WSe$_2$/NbSe$_2$ interface.

An overview of the electronic properties of similar interfaces, between monolayer WSe$_2$ atop many layer NbSe$_2$, is shown in Figure~\ref{Fig2}. In contrast to previous studies, we find that charge transfer from the WSe$_2$ flake to the bulk NbSe$_2$ substrate creates a semimetallic ground state with a negative Schottky barrier height of approximately -30~meV, thus indicating a perfect ohmic interface. Remarkably, the repositioning of the Fermi level to between the WSe$_2$ lVBM at K and $\Gamma$, results in a surface localised electronic structure with only momentum-separated spin-polarised band features. 

Band dispersions along the $\overline{\text{K}}$-$\overline{\Gamma}$-$\overline{\text{K}}$ and $\overline{\text{K}}$-$\overline{\text{M}}$-$\overline{\text{K}}$ directions in Fig.~\ref{Fig2}(d) and (e), respectively, show the full valence band structure of this charge doped system. With comparison to a free-standing ML-WSe$_2$ case (Fig.~\ref{Fig2}(c)), while the $d_{z^2}$-derived, lVBM at $\Gamma$ is significantly broadened, the remainder of the electronic structure, derived from in-plane orbitals with negligible out-of-plane hopping, is relatively unchanged other than the $\Delta E\approx$0.85~eV shift towards the Fermi level. This is calculated from the energy difference between the deeper lying K-point valley between the data in Fig.~\ref{Fig1} and Fig~\ref{Fig2}. 

The corresponding Fermi surface of this hybrid system is shown in Fig.~\ref{Fig2}(f). 
The bands at the K points derive from the lower binding energy branch of the spin-orbit split valleys in WSe$_2$, and the diffuse spectral weight at $\Gamma$ originates from the spectral tail of the broadened WSe$_2$ $\Gamma$-lVBM. There are also larger features with low spectral weight centered at both the $\Gamma$ and M points which we attribute to be signals directly from the substrate~\cite{borisenko_two_2009,bawden_spin_2015}. We note that due to the surface sensitivity of ARPES necessitating unimpeded access to sample regions of interest, characterizing the electronic structure of a semi metallic phase of WSe$_2$ without interfacing to NbSe$_2$ would require levels of electrostatic doping beyond that possible either with e.g. alkali metal deposition~\cite{riley_negative_2015} or through back gating exfoliated TMD flakes~\cite{nguyen_visualizing_2019}.

In Fig.~\ref{Fig3}, the contrasting dimensionalities of in- and out-of-plane orbital-derived bands (and therefore between the K and $\Gamma$ lVBM, respectively) are shown through photon energy dependent ARPES which changes  the effective $k_z$ plane probed in photoemission as well as the electron mean free path~\cite{Dam2004}. $\overline{\text{K}}$-$\overline{\Gamma}$-$\overline{\text{K}}$ band dispersions and Fermi surface maps for 121 and 150~eV photons, shown in Fig.~\ref{Fig3}(a,e) and~\ref{Fig3}(b,f) respectively, show increased spectral weight of bands imaged directly from the NbSe$_2$ substrate relative to the datasets in Fig.~\ref{Fig2} ($h\nu$=91.5~eV) due to the increased effective transparency of the top WSe$_2$ layer. This enables both direct confirmation of the preserved underlying NbSe$_2$ band structure and provides clear visualization of the 24$^{\circ}$ relative twist angle between substrate and the WSe$_2$ top layer. Band hybridizations driven by Moir\'e periodicity of lattice mismatched 2H-TMD systems are most evident for very small twist angles~\cite{devakul_magic_2021, guo_superconductivity_2024}, and is therefore likely absent here. In addition, the low-symmetry alignment between the two system components strongly suggests that the observed charge transfer does not require a specific lattice mismatching between components, and therefore the physics on show here can be exploited without complex assembly. 

In Fig.~\ref{Fig3}(d), the $k_z$ dependence of bands at the $\overline{\Gamma}$ point is shown. There are four features within this energy window labeled from low to high binding energy: The shallowest band closest to the Fermi level ($\#$1 in Fig.~\ref{Fig3}(d)) is the lVBM, which has become significantly broadened relative to the equivalent band in the free standing case (Fig.~\ref{Fig1}(f) and ~\ref{Fig2}(c)). Its dispersive nature (most clearly resolved in the high $k_z$ regime) evidences a delocalization along the $c$-axis into the bulk, thus permitting momentum dependence in the out-of-plane, $k_z$ direction. We attribute this broadening to be due to interactions of this W $d_{z^2}$-derived band with other out-of-plane orbital-derived bands which contribute to the near-$E_{\text{F}}$ band structure at $\overline{\Gamma}$ in the substrate (Nb $d_{z^2}$ and Se $p_z$). The other three bands are two dimensional: The deepest bands at approx. -1.4 and 1.7~eV ($\#$3 and $\#$4 in Fig.~\ref{Fig3}(d)) are mixed character Se-$p_{xy}$- and W $d_{xz,yz}$-derived bands which are unaffected by the details of the substrate due to the negligible $c-$axis hopping from in-plane orbitals~\cite{chhowalla_chemistry_2013, riley_direct_2014, bahramy_ubiquitous_2018}. The remaining band is a sharp 2D state at -0.8~eV without a clear analogue in the non-interacting limit. While this band could be mistaken for a second $k_z$ sub-band within the $d_{z^2}$ manifold, present within bilayer WSe$_2$, its two-dimensionality relative to the lower-energy band and the unambiguous photoluminsence signal (Fig.~\ref{Fig1}(b)) mean that this is unlikely. Instead, we attribute this to be a signature of the topological surface state predicted in bulk 2H-NbSe$_2$ at this energy and momentum~\cite{bahramy_ubiquitous_2018}. 

Figure~\ref{Fig3}(c) shows an equivalent $k_z$ dispersion for the $\overline{\text{K}}$ point: In contrast to the $\overline{\Gamma}$-lVBM, the pair of SOC split bands at $\overline{\text{K}}$ remain sharp, two-dimensional and therefore monolayer-like as a function of $k_z$, in line with the W $d_{xy,~x^2-y^2}$ orbitals from which they derive. The contrasting dimensionalities of the lVBM is further seen in the $k_y-k_z$ Fermi contour shown in Fig.~\ref{Fig3}(g), where bands near $\overline{\Gamma}$ and $\overline{\text{K}}$ are diffuse and well-defined, respectively.

Together, this evidences that  the states at $\overline{\text{K}}$ remain localized to the WSe$_2$ toplayer, and can therefore be considered surface states of the hybrid bulk system. There has been significant effort in characterizing materials with large atomic spin-orbit coupling which are capable of hosting Rashba split states either at the surface (due to the surface potential step) or within the bulk (due to non-centrosymmetric crystal structures) with a large band separation~\cite{zhai_transient_2017, sunko_maximal_2017, ishizaka_giant_2011, bian_origin_2013, bianchi_robust_2012, feng_giant_2025,picozzi_ferroelectric_2014}. The surface states of this hybrid system can be considered a particularly extreme example of this, with up and down spin species separated by the K-K' separation of the BZ ($\frac{4\pi}{3a} \approx$ 1.3~\invA). The momentum separation of these states should provide significant resilience against inter-band scattering, similar to that found in metallic surface states of topological insulators (TIs)~\cite{hasan_colloquium_2010}. 

\begin{figure}[t]
	\centering	\includegraphics[width=\columnwidth]{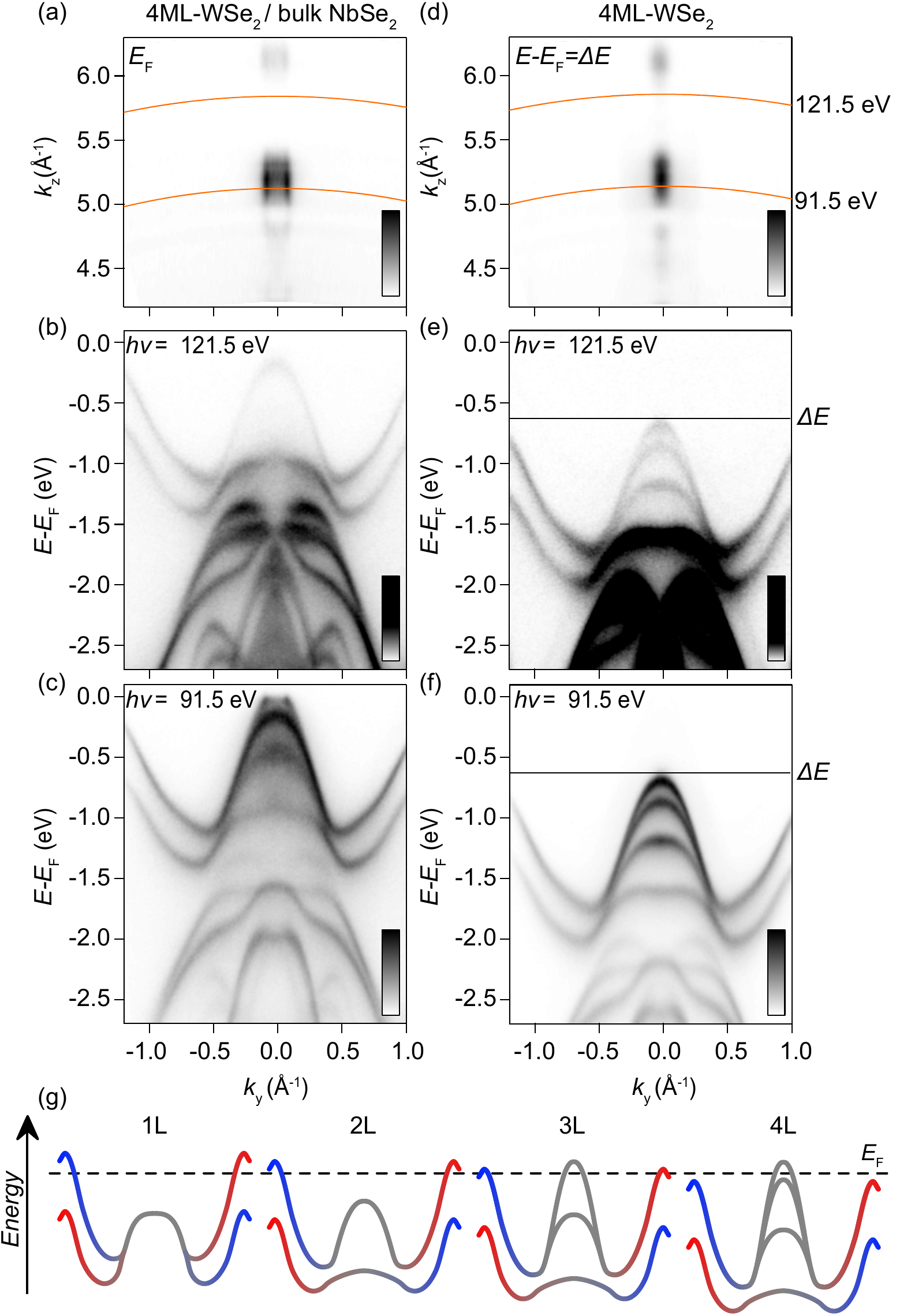}
	\caption{\label{Fig4} \textbf{Singular Fermi surface in 4ML-WSe$_2$/bulk NbSe$_2$.} (a) Fermi level $k_y$-$k_z$ contour showing the dimensionality of the Fermi pocket at $\Gamma$. Orange lines indicate the positioning of band dispersions presented in (b-c). (b-c) Band dispersions close to the K-$\Gamma$-K axis, taken with 121.5 (b) and 91.5~eV (c) photons. (d-f) Equivalent datasets for 4ML-WSe$_2$/gr. The $k_y$-$k_z$ contour in (d) is presented for $E-E_{\text{F}} = \Delta E$, the calculated charge transfer when interfaced with bulk NbSe$_2$. (g) Schematic demonstrating the likely Fermi level placement for 1-4 ML-WSe$_2$/bulk NbSe$_2$.} 
\end{figure}

To finalize, we demonstrate how the broader physics of 2H-WSe$_2$ is maintained despite the significant charge doping. The $\Gamma$ lVBM in free-standing WSe$_2$ forms a dispersive band continuum along $k_z$ in the bulk limit due to non-negligible out-of-plane hopping of the $d_{z^2}$-orbitals from which it derives. For intermediate thicknesses, it forms an array of energetically-separated, doubly-degenerate $k_z$ subbands; the number of which matches the total thickness in monolayers. Since the details of K point lVBM are relatively unaffected by the surrounding layers, the $\Gamma$-lVBM  becomes the global valence band maximum for 3+ ML ~\cite{latzke_electronic_2015, kim_determination_2016, alarab_k_2023, lefevre_two_2024, chang_thickness_2014}. In Fig.~\ref{Fig4}, we show how this $c-$axis delocalization of out-of-plane orbital derived electrons within the WSe$_2$ top-layers drastically alters the fermiology of interfaced systems. In Fig.~\ref{Fig4}(a-f), ARPES spectra are compared for 4ML-WSe$_2$, where the global valence band maximum is at the $\Gamma$ point~\cite{mak_atomically_2010, sun_indirect_2016, ernandes_indirect_2021} and a 4ML-WSe$_2$/bulk NbSe$_2$ system. In analogy to the monolayer case in Figure~\ref{Fig2}, the charge transfer acts to shift the bands towards the Fermi level by $\Delta E \approx$ 0.63~eV, as  calculated from the shift of the 2D state at $\Gamma$ and $\sim$ -2.2~eV in pristine 4ML-WSe$_2$. In this scenario, the global-VBM of 4ML-WSe$_2$ at $\Gamma$ creates a singular Fermi surface. As the four quantized bands at $\overline{\Gamma}$ sample the continuous $k_z$-dispersive global VBM of bulk WSe$_2$, the Fermi pocket is not visible at all photon energies, instead exhibiting periodic spectral weight in $k_z$ (Fig.~\ref{Fig4}(a and d)). The valence band top is therefore visible and absent for 91.5~eV and 121.5 eV photons, respectively, as shown in Fig.~\ref{Fig4}(b-c). We note that the spacing and appearance of the four discrete W-d$_{z^2}$ subbands at $\Gamma$ are slightly altered relative to free-standing 4ML-WSe$_2$ (Fig.~\ref{Fig4}(d-f)), again likely due to interactions with the out-of-plane Nb $d$ and Se $p$-orbital-derived bands of the bulk substrate. 

The energy shift of the bands in the 4ML scenario is smaller than observed for the monolayer limit ($\approx$ 0.85~eV), but the resulting Luttinger count is  similar ($\approx 10^{-13}$ cm$^{-2}$ in both cases when assuming that Fermi surfaces formed in the ML and 4L scenarios have spin degeneracies of 1 and 2, respectively~\cite{luttinger_1960}). This reduced potential difference and constant carrier density are consistent with the increased thickness of the WSe$_2$ top layer, and therefore the increased separation of displaced charge. 

As schematized in Fig.~\ref{Fig4}(g), one should therefore expect that the 2 and 3L scenarios produce the intermediate scenarios: the gradual shift of the global VBM from K to $\Gamma$ with the increasing film thickness should result in distinct Fermi surfaces from those observed in our systems, and there may exist a thickness where both Fermi surfaces exist simultaneously.  A similar $\Gamma$-centred fermiology was achieved recently by interfacing few-layer WSe$_2$  with few-layer NbSe$_2$, demonstrating how these observed ohmic interfaces do not require a bulk component, though the Fermi surface is likely always at $\Gamma$ for pairs of few-layer flakes~\cite{clark_dimensional_2025}.


\section{\label{sec:conc}CONCLUSION\protect\\ }

In conclusion, we have demonstrated that matching thin flakes of WSe$_2$ to many-layer NbSe$_2$ single crystals creates ohmic $p$-type contacts, and the momenta and spin-degeneracy of the charge carriers at the interface can be tuned with thickness of the WSe$_2$ overlay. 

For the case of monolayer WSe$_2$, the Fermi level is shifted between the local band maxima at the $\Gamma$ and K points. The oppositely spin-polarised bands at K' and K remain localized to the WSe$_2$ monolayer due to their in-plane orbital makeup, whereas the band at $\Gamma$ becomes resonant with out-of-plane orbitals from NbSe$_2$, and hence delocalised into the bulk. The VBM at K/K' are  thus the sole bands localized to the surface, and therefore the hybrid system becomes analogous to a bulk solid wherein Rashba-split surface states cross the Fermi level, but with out-of-plane spin polarisation and a much increased momentum separation.

The semimetallic equilibrium state found here is likely further tunable through gating, and the relative size of the spin-polarised hole pockets at K and K' could be potentially changed with e.g. uniaxial strain or applied electric fields. Similar charge transfer mechanics should apply to other pairings of metallic 2H-TMDs with 2H-TMD semiconductors and their lattice mismatched few-layer devices, together producing a large parameter space within which these novel semimetallic phases can be tuned. 

It is noteworthy that 2H-NbSe$_2$ possesses competing charge density wave (CDW) and superconducting orders, with transition temperatures of $\approx$ 34 and 7~K, respectively in the bulk limit, but are each strongly thickness dependent~\cite{xi_strongly_2015, dvir_spectroscopy_2018,hamill_two_2021, xi_ising_2016}. While the twist angle between the WSe$_2$ monolayer and bulk NbSe$_2$ is large in this proof-of-principle example, smaller twist angles could lead to a pronounced Moiré periodicity, which could be sensitive to the CDW reconstruction of the substrate~\cite{yoshizawa_visualization_2024}.  More excitingly, the numerous superconducting phases of NbSe$_2$ could lead to odd-parity pairing between the spin-polarised Fermi pockets, a scenario which is thought one pathway towards topological superconductivity~\cite{sato_topological_2017}. More generally, this work bolsters the collective electronic properties of interfaced TMD systems~\cite{shao_strong_2016, he_all_2025, vu_synthesis_2024, clark_dimensional_2025, devakul_magic_2021, guo_superconductivity_2024,fang_localization_2023}, while providing the direct evidence for perfect ohmic interfaces with negative Schottky barriers between a TMD semiconductor and metal; a result of technological importance for the ultimate optimization and miniaturization of electronic devices. 

\section{\label{sec:methods}METHODS\protect\\ }

TMD flakes were exfoliated onto PDMS, and their thicknesses were verified through photoluminesence spectroscopy (e.g. Fig.~\ref{Fig1}(b)) and by comparing optical images and photoemission spectra to those of known references. High-quality single crystals of NbSe$_2$ were cleaved in an Ar-atmosphere using a standard top post method. The ML-WSe$_2$/bulk NbSe$_2$ data originate from the a WSe$_2$ monolayer transferred from PDMS onto the bulk NbSe$_2$ cleaved surface. The remaining spectra originate from  WSe$_2$, or WSe$_2$ and NbSe$_2$ flakes transferred onto graphite/Si substrates. All samples were rinsed in diisopropanol, isopropanol and ethanol, and annealed in UHV prior to photoemission experiments. Photoemission data were acquired using either a nano-ESCA system with a helium lamp (NanoESCA III at FMMA (ROR: 04z91ja70)~\cite{FlindersNanoESCA}) at room temperature (Fig.~\ref{Fig1}) or by nano-focussed angle-resolved photoemission (nano-ARPES) performed at the MAESTRO beamline of the Advanced Light Source (LBNL) at cryogenic temperatures of $\approx$25~K with $p$-polarised photons of energies between 60eV and 150eV. Mapping from photon energy to $k_z$ was done using free-electron final state assumption~\cite{Dam2004}, using an inner potential of 13~eV, matching that found for bulk WSe$_2$~\cite{riley_direct_2014}.

\bibliographystyle{apsrev4-1}

\bibliography{bib.bib}

\begin{thebibliography}{60}%
\makeatletter
\providecommand \@ifxundefined [1]{%
 \@ifx{#1\undefined}
}%
\providecommand \@ifnum [1]{%
 \ifnum #1\expandafter \@firstoftwo
 \else \expandafter \@secondoftwo
 \fi
}%
\providecommand \@ifx [1]{%
 \ifx #1\expandafter \@firstoftwo
 \else \expandafter \@secondoftwo
 \fi
}%
\providecommand \natexlab [1]{#1}%
\providecommand \enquote  [1]{``#1''}%
\providecommand \bibnamefont  [1]{#1}%
\providecommand \bibfnamefont [1]{#1}%
\providecommand \citenamefont [1]{#1}%
\providecommand \href@noop [0]{\@secondoftwo}%
\providecommand \href [0]{\begingroup \@sanitize@url \@href}%
\providecommand \@href[1]{\@@startlink{#1}\@@href}%
\providecommand \@@href[1]{\endgroup#1\@@endlink}%
\providecommand \@sanitize@url [0]{\catcode `\\12\catcode `\$12\catcode `\&12\catcode `\#12\catcode `\^12\catcode `\_12\catcode `\%12\relax}%
\providecommand \@@startlink[1]{}%
\providecommand \@@endlink[0]{}%
\providecommand \url  [0]{\begingroup\@sanitize@url \@url }%
\providecommand \@url [1]{\endgroup\@href {#1}{\urlprefix }}%
\providecommand \urlprefix  [0]{URL }%
\providecommand \Eprint [0]{\href }%
\providecommand \doibase [0]{http://dx.doi.org/}%
\providecommand \selectlanguage [0]{\@gobble}%
\providecommand \bibinfo  [0]{\@secondoftwo}%
\providecommand \bibfield  [0]{\@secondoftwo}%
\providecommand \translation [1]{[#1]}%
\providecommand \BibitemOpen [0]{}%
\providecommand \bibitemStop [0]{}%
\providecommand \bibitemNoStop [0]{.\EOS\space}%
\providecommand \EOS [0]{\spacefactor3000\relax}%
\providecommand \BibitemShut  [1]{\csname bibitem#1\endcsname}%
\let\auto@bib@innerbib\@empty
\bibitem [{\citenamefont {Patoary}\ \emph {et~al.}(2023)\citenamefont {Patoary}, \citenamefont {Xie}, \citenamefont {Zhou}, \citenamefont {Al~Mamun}, \citenamefont {Sayyad}, \citenamefont {Tongay},\ and\ \citenamefont {Esqueda}}]{patoary_improvements_2023}%
  \BibitemOpen
  \bibfield  {author} {\bibinfo {author} {\bibfnamefont {N.~H.}\ \bibnamefont {Patoary}}, \bibinfo {author} {\bibfnamefont {J.}~\bibnamefont {Xie}}, \bibinfo {author} {\bibfnamefont {G.}~\bibnamefont {Zhou}}, \bibinfo {author} {\bibfnamefont {F.}~\bibnamefont {Al~Mamun}}, \bibinfo {author} {\bibfnamefont {M.}~\bibnamefont {Sayyad}}, \bibinfo {author} {\bibfnamefont {S.}~\bibnamefont {Tongay}}, \ and\ \bibinfo {author} {\bibfnamefont {I.~S.}\ \bibnamefont {Esqueda}},\ }\href {\doibase 10.1038/s41598-023-30317-4} {\bibfield  {journal} {\bibinfo  {journal} {Scientific Reports}\ }\textbf {\bibinfo {volume} {13}},\ \bibinfo {pages} {3304} (\bibinfo {year} {2023})}\BibitemShut {NoStop}%
\bibitem [{\citenamefont {Park}\ \emph {et~al.}(2025)\citenamefont {Park}, \citenamefont {Shin}, \citenamefont {Ali}, \citenamefont {Choi}, \citenamefont {Kim}, \citenamefont {Kang}, \citenamefont {Choi},\ and\ \citenamefont {Yoo}}]{park_p_2025}%
  \BibitemOpen
  \bibfield  {author} {\bibinfo {author} {\bibfnamefont {H.}~\bibnamefont {Park}}, \bibinfo {author} {\bibfnamefont {H.}~\bibnamefont {Shin}}, \bibinfo {author} {\bibfnamefont {N.}~\bibnamefont {Ali}}, \bibinfo {author} {\bibfnamefont {H.}~\bibnamefont {Choi}}, \bibinfo {author} {\bibfnamefont {B.~S.~Y.}\ \bibnamefont {Kim}}, \bibinfo {author} {\bibfnamefont {B.}~\bibnamefont {Kang}}, \bibinfo {author} {\bibfnamefont {M.~S.}\ \bibnamefont {Choi}}, \ and\ \bibinfo {author} {\bibfnamefont {W.~J.}\ \bibnamefont {Yoo}},\ }\href {\doibase 10.1021/acs.nanolett.4c05136} {\bibfield  {journal} {\bibinfo  {journal} {Nano Letters}\ }\textbf {\bibinfo {volume} {25}},\ \bibinfo {pages} {368} (\bibinfo {year} {2025})}\BibitemShut {NoStop}%
\bibitem [{\citenamefont {Sata}\ \emph {et~al.}(2017)\citenamefont {Sata}, \citenamefont {Moriya}, \citenamefont {Masubuchi}, \citenamefont {Watanabe}, \citenamefont {Taniguchi},\ and\ \citenamefont {Machida}}]{sata_n_2017}%
  \BibitemOpen
  \bibfield  {author} {\bibinfo {author} {\bibfnamefont {Y.}~\bibnamefont {Sata}}, \bibinfo {author} {\bibfnamefont {R.}~\bibnamefont {Moriya}}, \bibinfo {author} {\bibfnamefont {S.}~\bibnamefont {Masubuchi}}, \bibinfo {author} {\bibfnamefont {K.}~\bibnamefont {Watanabe}}, \bibinfo {author} {\bibfnamefont {T.}~\bibnamefont {Taniguchi}}, \ and\ \bibinfo {author} {\bibfnamefont {T.}~\bibnamefont {Machida}},\ }\href {\doibase 10.7567/JJAP.56.04CK09} {\bibfield  {journal} {\bibinfo  {journal} {Japanese Journal of Applied Physics}\ }\textbf {\bibinfo {volume} {56}},\ \bibinfo {pages} {04CK09} (\bibinfo {year} {2017})}\BibitemShut {NoStop}%
\bibitem [{\citenamefont {Hu}\ \emph {et~al.}(2025)\citenamefont {Hu}, \citenamefont {Wang}, \citenamefont {Pan}, \citenamefont {Chen}, \citenamefont {Han},\ and\ \citenamefont {Wang}}]{hu_interface_2025}%
  \BibitemOpen
  \bibfield  {author} {\bibinfo {author} {\bibfnamefont {H.}~\bibnamefont {Hu}}, \bibinfo {author} {\bibfnamefont {Z.}~\bibnamefont {Wang}}, \bibinfo {author} {\bibfnamefont {M.}~\bibnamefont {Pan}}, \bibinfo {author} {\bibfnamefont {Y.}~\bibnamefont {Chen}}, \bibinfo {author} {\bibfnamefont {Y.}~\bibnamefont {Han}}, \ and\ \bibinfo {author} {\bibfnamefont {J.}~\bibnamefont {Wang}},\ }\href {\doibase https://doi.org/10.1002/advs.202500226} {\bibfield  {journal} {\bibinfo  {journal} {Advanced Science}\ }\textbf {\bibinfo {volume} {12}},\ \bibinfo {pages} {2500226} (\bibinfo {year} {2025})}\BibitemShut {NoStop}%
\bibitem [{\citenamefont {Datta}\ and\ \citenamefont {Das}(1990)}]{datta_electronic_1990}%
  \BibitemOpen
  \bibfield  {author} {\bibinfo {author} {\bibfnamefont {S.}~\bibnamefont {Datta}}\ and\ \bibinfo {author} {\bibfnamefont {B.}~\bibnamefont {Das}},\ }\href {\doibase 10.1063/1.102730} {\bibfield  {journal} {\bibinfo  {journal} {Applied Physics Letters}\ }\textbf {\bibinfo {volume} {56}},\ \bibinfo {pages} {665} (\bibinfo {year} {1990})},\ \Eprint {http://arxiv.org/abs/https://doi.org/10.1063/1.102730} {https://doi.org/10.1063/1.102730} \BibitemShut {NoStop}%
\bibitem [{\citenamefont {Gor'kov}\ and\ \citenamefont {Rashba}(2001)}]{gorkov_superconducting_2001}%
  \BibitemOpen
  \bibfield  {author} {\bibinfo {author} {\bibfnamefont {L.~P.}\ \bibnamefont {Gor'kov}}\ and\ \bibinfo {author} {\bibfnamefont {E.~I.}\ \bibnamefont {Rashba}},\ }\href {\doibase 10.1103/PhysRevLett.87.037004} {\bibfield  {journal} {\bibinfo  {journal} {Phys. Rev. Lett.}\ }\textbf {\bibinfo {volume} {87}},\ \bibinfo {pages} {037004} (\bibinfo {year} {2001})}\BibitemShut {NoStop}%
\bibitem [{\citenamefont {Lesne}\ \emph {et~al.}(2016)\citenamefont {Lesne}, \citenamefont {Fu}, \citenamefont {Oyarzun}, \citenamefont {Rojas-S{\'a}nchez}, \citenamefont {Vaz}, \citenamefont {Naganuma}, \citenamefont {Sicoli}, \citenamefont {Attan{\'e}}, \citenamefont {Jamet}, \citenamefont {Jacquet}, \citenamefont {George}, \citenamefont {Barth{\'e}l{\'e}my}, \citenamefont {Jaffr{\`e}s}, \citenamefont {Fert}, \citenamefont {Bibes},\ and\ \citenamefont {Vila}}]{lesne_highly_2016}%
  \BibitemOpen
  \bibfield  {author} {\bibinfo {author} {\bibfnamefont {E.}~\bibnamefont {Lesne}}, \bibinfo {author} {\bibfnamefont {Y.}~\bibnamefont {Fu}}, \bibinfo {author} {\bibfnamefont {S.}~\bibnamefont {Oyarzun}}, \bibinfo {author} {\bibfnamefont {J.~C.}\ \bibnamefont {Rojas-S{\'a}nchez}}, \bibinfo {author} {\bibfnamefont {D.~C.}\ \bibnamefont {Vaz}}, \bibinfo {author} {\bibfnamefont {H.}~\bibnamefont {Naganuma}}, \bibinfo {author} {\bibfnamefont {G.}~\bibnamefont {Sicoli}}, \bibinfo {author} {\bibfnamefont {J.~P.}\ \bibnamefont {Attan{\'e}}}, \bibinfo {author} {\bibfnamefont {M.}~\bibnamefont {Jamet}}, \bibinfo {author} {\bibfnamefont {E.}~\bibnamefont {Jacquet}}, \bibinfo {author} {\bibfnamefont {J.~M.}\ \bibnamefont {George}}, \bibinfo {author} {\bibfnamefont {A.}~\bibnamefont {Barth{\'e}l{\'e}my}}, \bibinfo {author} {\bibfnamefont {H.}~\bibnamefont {Jaffr{\`e}s}}, \bibinfo {author} {\bibfnamefont {A.}~\bibnamefont {Fert}}, \bibinfo {author} {\bibfnamefont {M.}~\bibnamefont {Bibes}}, \ and\ \bibinfo {author}
  {\bibfnamefont {L.}~\bibnamefont {Vila}},\ }\href {\doibase 10.1038/nmat4726} {\bibfield  {journal} {\bibinfo  {journal} {Nature Materials}\ }\textbf {\bibinfo {volume} {15}},\ \bibinfo {pages} {1261} (\bibinfo {year} {2016})}\BibitemShut {NoStop}%
\bibitem [{\citenamefont {Manchon}\ \emph {et~al.}(2015)\citenamefont {Manchon}, \citenamefont {Koo}, \citenamefont {Nitta}, \citenamefont {Frolov},\ and\ \citenamefont {Duine}}]{manchon_new_2015}%
  \BibitemOpen
  \bibfield  {author} {\bibinfo {author} {\bibfnamefont {A.}~\bibnamefont {Manchon}}, \bibinfo {author} {\bibfnamefont {H.~C.}\ \bibnamefont {Koo}}, \bibinfo {author} {\bibfnamefont {J.}~\bibnamefont {Nitta}}, \bibinfo {author} {\bibfnamefont {S.~M.}\ \bibnamefont {Frolov}}, \ and\ \bibinfo {author} {\bibfnamefont {R.~A.}\ \bibnamefont {Duine}},\ }\href {\doibase 10.1038/nmat4360} {\bibfield  {journal} {\bibinfo  {journal} {Nature Materials}\ }\textbf {\bibinfo {volume} {14}},\ \bibinfo {pages} {871} (\bibinfo {year} {2015})}\BibitemShut {NoStop}%
\bibitem [{\citenamefont {Hirohata}\ \emph {et~al.}(2020)\citenamefont {Hirohata}, \citenamefont {Yamada}, \citenamefont {Nakatani}, \citenamefont {Prejbeanu}, \citenamefont {Diény}, \citenamefont {Pirro},\ and\ \citenamefont {Hillebrands}}]{hirohata_review_2020}%
  \BibitemOpen
  \bibfield  {author} {\bibinfo {author} {\bibfnamefont {A.}~\bibnamefont {Hirohata}}, \bibinfo {author} {\bibfnamefont {K.}~\bibnamefont {Yamada}}, \bibinfo {author} {\bibfnamefont {Y.}~\bibnamefont {Nakatani}}, \bibinfo {author} {\bibfnamefont {I.-L.}\ \bibnamefont {Prejbeanu}}, \bibinfo {author} {\bibfnamefont {B.}~\bibnamefont {Diény}}, \bibinfo {author} {\bibfnamefont {P.}~\bibnamefont {Pirro}}, \ and\ \bibinfo {author} {\bibfnamefont {B.}~\bibnamefont {Hillebrands}},\ }\href {\doibase https://doi.org/10.1016/j.jmmm.2020.166711} {\bibfield  {journal} {\bibinfo  {journal} {Journal of Magnetism and Magnetic Materials}\ }\textbf {\bibinfo {volume} {509}},\ \bibinfo {pages} {166711} (\bibinfo {year} {2020})}\BibitemShut {NoStop}%
\bibitem [{\citenamefont {Breunig}\ and\ \citenamefont {Ando}(2022)}]{breunig_opportunities_2022}%
  \BibitemOpen
  \bibfield  {author} {\bibinfo {author} {\bibfnamefont {O.}~\bibnamefont {Breunig}}\ and\ \bibinfo {author} {\bibfnamefont {Y.}~\bibnamefont {Ando}},\ }\href {\doibase 10.1038/s42254-021-00402-6} {\bibfield  {journal} {\bibinfo  {journal} {Nature Reviews Physics}\ }\textbf {\bibinfo {volume} {4}},\ \bibinfo {pages} {184} (\bibinfo {year} {2022})}\BibitemShut {NoStop}%
\bibitem [{\citenamefont {Bychkov}\ and\ \citenamefont {Rashba}(1984)}]{Rashba84}%
  \BibitemOpen
  \bibfield  {author} {\bibinfo {author} {\bibfnamefont {Y.~A.}\ \bibnamefont {Bychkov}}\ and\ \bibinfo {author} {\bibfnamefont {E.~I.}\ \bibnamefont {Rashba}},\ }\bibfield  {booktitle} {\emph {\bibinfo {booktitle} {Journal of Physics C: Solid State Physics}},\ }\href {\doibase 10.1088/0022-3719/17/33/015} {\ \textbf {\bibinfo {volume} {17}},\ \bibinfo {pages} {6039} (\bibinfo {year} {1984})}\BibitemShut {NoStop}%
\bibitem [{\citenamefont {Ogawa}\ \emph {et~al.}(2014)\citenamefont {Ogawa}, \citenamefont {Bahramy}, \citenamefont {Kaneko},\ and\ \citenamefont {Tokura}}]{ogowa_photocontrol_2014}%
  \BibitemOpen
  \bibfield  {author} {\bibinfo {author} {\bibfnamefont {N.}~\bibnamefont {Ogawa}}, \bibinfo {author} {\bibfnamefont {M.~S.}\ \bibnamefont {Bahramy}}, \bibinfo {author} {\bibfnamefont {Y.}~\bibnamefont {Kaneko}}, \ and\ \bibinfo {author} {\bibfnamefont {Y.}~\bibnamefont {Tokura}},\ }\href {\doibase 10.1103/PhysRevB.90.125122} {\bibfield  {journal} {\bibinfo  {journal} {Phys. Rev. B}\ }\textbf {\bibinfo {volume} {90}},\ \bibinfo {pages} {125122} (\bibinfo {year} {2014})}\BibitemShut {NoStop}%
\bibitem [{\citenamefont {S{\'a}nchez}\ \emph {et~al.}(2013)\citenamefont {S{\'a}nchez}, \citenamefont {Vila}, \citenamefont {Desfonds}, \citenamefont {Gambarelli}, \citenamefont {Attan{\'e}}, \citenamefont {De~Teresa}, \citenamefont {Mag{\'e}n},\ and\ \citenamefont {Fert}}]{sanchez_spin_2013}%
  \BibitemOpen
  \bibfield  {author} {\bibinfo {author} {\bibfnamefont {J.~C.~R.}\ \bibnamefont {S{\'a}nchez}}, \bibinfo {author} {\bibfnamefont {L.}~\bibnamefont {Vila}}, \bibinfo {author} {\bibfnamefont {G.}~\bibnamefont {Desfonds}}, \bibinfo {author} {\bibfnamefont {S.}~\bibnamefont {Gambarelli}}, \bibinfo {author} {\bibfnamefont {J.~P.}\ \bibnamefont {Attan{\'e}}}, \bibinfo {author} {\bibfnamefont {J.~M.}\ \bibnamefont {De~Teresa}}, \bibinfo {author} {\bibfnamefont {C.}~\bibnamefont {Mag{\'e}n}}, \ and\ \bibinfo {author} {\bibfnamefont {A.}~\bibnamefont {Fert}},\ }\href {\doibase 10.1038/ncomms3944} {\bibfield  {journal} {\bibinfo  {journal} {Nature Communications}\ }\textbf {\bibinfo {volume} {4}},\ \bibinfo {pages} {2944} (\bibinfo {year} {2013})}\BibitemShut {NoStop}%
\bibitem [{\citenamefont {Jungfleisch}\ \emph {et~al.}(2018)\citenamefont {Jungfleisch}, \citenamefont {Zhang}, \citenamefont {Zhang}, \citenamefont {Pearson}, \citenamefont {Schaller}, \citenamefont {Wen},\ and\ \citenamefont {Hoffmann}}]{jungfleisch_control_2018}%
  \BibitemOpen
  \bibfield  {author} {\bibinfo {author} {\bibfnamefont {M.~B.}\ \bibnamefont {Jungfleisch}}, \bibinfo {author} {\bibfnamefont {Q.}~\bibnamefont {Zhang}}, \bibinfo {author} {\bibfnamefont {W.}~\bibnamefont {Zhang}}, \bibinfo {author} {\bibfnamefont {J.~E.}\ \bibnamefont {Pearson}}, \bibinfo {author} {\bibfnamefont {R.~D.}\ \bibnamefont {Schaller}}, \bibinfo {author} {\bibfnamefont {H.}~\bibnamefont {Wen}}, \ and\ \bibinfo {author} {\bibfnamefont {A.}~\bibnamefont {Hoffmann}},\ }\href {\doibase 10.1103/PhysRevLett.120.207207} {\bibfield  {journal} {\bibinfo  {journal} {Phys. Rev. Lett.}\ }\textbf {\bibinfo {volume} {120}},\ \bibinfo {pages} {207207} (\bibinfo {year} {2018})}\BibitemShut {NoStop}%
\bibitem [{\citenamefont {Varotto}\ \emph {et~al.}(2021)\citenamefont {Varotto}, \citenamefont {Nessi}, \citenamefont {Cecchi}, \citenamefont {S{\l}awi{\'n}ska}, \citenamefont {No{\"e}l}, \citenamefont {Petr{\`o}}, \citenamefont {Fagiani}, \citenamefont {Novati}, \citenamefont {Cantoni}, \citenamefont {Petti}, \citenamefont {Albisetti}, \citenamefont {Costa}, \citenamefont {Calarco}, \citenamefont {Buongiorno~Nardelli}, \citenamefont {Bibes}, \citenamefont {Picozzi}, \citenamefont {Attan{\'e}}, \citenamefont {Vila}, \citenamefont {Bertacco},\ and\ \citenamefont {Rinaldi}}]{varotto_room_2021}%
  \BibitemOpen
  \bibfield  {author} {\bibinfo {author} {\bibfnamefont {S.}~\bibnamefont {Varotto}}, \bibinfo {author} {\bibfnamefont {L.}~\bibnamefont {Nessi}}, \bibinfo {author} {\bibfnamefont {S.}~\bibnamefont {Cecchi}}, \bibinfo {author} {\bibfnamefont {J.}~\bibnamefont {S{\l}awi{\'n}ska}}, \bibinfo {author} {\bibfnamefont {P.}~\bibnamefont {No{\"e}l}}, \bibinfo {author} {\bibfnamefont {S.}~\bibnamefont {Petr{\`o}}}, \bibinfo {author} {\bibfnamefont {F.}~\bibnamefont {Fagiani}}, \bibinfo {author} {\bibfnamefont {A.}~\bibnamefont {Novati}}, \bibinfo {author} {\bibfnamefont {M.}~\bibnamefont {Cantoni}}, \bibinfo {author} {\bibfnamefont {D.}~\bibnamefont {Petti}}, \bibinfo {author} {\bibfnamefont {E.}~\bibnamefont {Albisetti}}, \bibinfo {author} {\bibfnamefont {M.}~\bibnamefont {Costa}}, \bibinfo {author} {\bibfnamefont {R.}~\bibnamefont {Calarco}}, \bibinfo {author} {\bibfnamefont {M.}~\bibnamefont {Buongiorno~Nardelli}}, \bibinfo {author} {\bibfnamefont {M.}~\bibnamefont {Bibes}}, \bibinfo {author} {\bibfnamefont
  {S.}~\bibnamefont {Picozzi}}, \bibinfo {author} {\bibfnamefont {J.-P.}\ \bibnamefont {Attan{\'e}}}, \bibinfo {author} {\bibfnamefont {L.}~\bibnamefont {Vila}}, \bibinfo {author} {\bibfnamefont {R.}~\bibnamefont {Bertacco}}, \ and\ \bibinfo {author} {\bibfnamefont {C.}~\bibnamefont {Rinaldi}},\ }\href {\doibase 10.1038/s41928-021-00653-2} {\bibfield  {journal} {\bibinfo  {journal} {Nature Electronics}\ }\textbf {\bibinfo {volume} {4}},\ \bibinfo {pages} {740} (\bibinfo {year} {2021})}\BibitemShut {NoStop}%
\bibitem [{\citenamefont {Rinaldi}\ \emph {et~al.}(2018)\citenamefont {Rinaldi}, \citenamefont {Varotto}, \citenamefont {Asa}, \citenamefont {S{\l}awi{\'n}ska}, \citenamefont {Fujii}, \citenamefont {Vinai}, \citenamefont {Cecchi}, \citenamefont {Di~Sante}, \citenamefont {Calarco}, \citenamefont {Vobornik}, \citenamefont {Panaccione}, \citenamefont {Picozzi},\ and\ \citenamefont {Bertacco}}]{rinaldi_ferroelectric_2018}%
  \BibitemOpen
  \bibfield  {author} {\bibinfo {author} {\bibfnamefont {C.}~\bibnamefont {Rinaldi}}, \bibinfo {author} {\bibfnamefont {S.}~\bibnamefont {Varotto}}, \bibinfo {author} {\bibfnamefont {M.}~\bibnamefont {Asa}}, \bibinfo {author} {\bibfnamefont {J.}~\bibnamefont {S{\l}awi{\'n}ska}}, \bibinfo {author} {\bibfnamefont {J.}~\bibnamefont {Fujii}}, \bibinfo {author} {\bibfnamefont {G.}~\bibnamefont {Vinai}}, \bibinfo {author} {\bibfnamefont {S.}~\bibnamefont {Cecchi}}, \bibinfo {author} {\bibfnamefont {D.}~\bibnamefont {Di~Sante}}, \bibinfo {author} {\bibfnamefont {R.}~\bibnamefont {Calarco}}, \bibinfo {author} {\bibfnamefont {I.}~\bibnamefont {Vobornik}}, \bibinfo {author} {\bibfnamefont {G.}~\bibnamefont {Panaccione}}, \bibinfo {author} {\bibfnamefont {S.}~\bibnamefont {Picozzi}}, \ and\ \bibinfo {author} {\bibfnamefont {R.}~\bibnamefont {Bertacco}},\ }\href {\doibase 10.1021/acs.nanolett.7b04829} {\bibfield  {journal} {\bibinfo  {journal} {Nano Letters}\ }\textbf {\bibinfo {volume} {18}},\ \bibinfo {pages} {2751}
  (\bibinfo {year} {2018})}\BibitemShut {NoStop}%
\bibitem [{\citenamefont {Xiao}\ \emph {et~al.}(2012)\citenamefont {Xiao}, \citenamefont {Liu}, \citenamefont {Feng}, \citenamefont {Xu},\ and\ \citenamefont {Yao}}]{xaio_coupled_2012}%
  \BibitemOpen
  \bibfield  {author} {\bibinfo {author} {\bibfnamefont {D.}~\bibnamefont {Xiao}}, \bibinfo {author} {\bibfnamefont {G.-B.}\ \bibnamefont {Liu}}, \bibinfo {author} {\bibfnamefont {W.}~\bibnamefont {Feng}}, \bibinfo {author} {\bibfnamefont {X.}~\bibnamefont {Xu}}, \ and\ \bibinfo {author} {\bibfnamefont {W.}~\bibnamefont {Yao}},\ }\href {\doibase 10.1103/PhysRevLett.108.196802} {\bibfield  {journal} {\bibinfo  {journal} {Phys. Rev. Lett.}\ }\textbf {\bibinfo {volume} {108}},\ \bibinfo {pages} {196802} (\bibinfo {year} {2012})}\BibitemShut {NoStop}%
\bibitem [{\citenamefont {Zeng}\ \emph {et~al.}(2012)\citenamefont {Zeng}, \citenamefont {Dai}, \citenamefont {Yao}, \citenamefont {Xiao},\ and\ \citenamefont {Cui}}]{zeng_valley_2012}%
  \BibitemOpen
  \bibfield  {author} {\bibinfo {author} {\bibfnamefont {H.}~\bibnamefont {Zeng}}, \bibinfo {author} {\bibfnamefont {J.}~\bibnamefont {Dai}}, \bibinfo {author} {\bibfnamefont {W.}~\bibnamefont {Yao}}, \bibinfo {author} {\bibfnamefont {D.}~\bibnamefont {Xiao}}, \ and\ \bibinfo {author} {\bibfnamefont {X.}~\bibnamefont {Cui}},\ }\href {\doibase 10.1038/nnano.2012.95} {\bibfield  {journal} {\bibinfo  {journal} {Nature Nanotechnology}\ }\textbf {\bibinfo {volume} {7}},\ \bibinfo {pages} {490} (\bibinfo {year} {2012})}\BibitemShut {NoStop}%
\bibitem [{\citenamefont {Mak}\ \emph {et~al.}(2012)\citenamefont {Mak}, \citenamefont {He}, \citenamefont {Shan},\ and\ \citenamefont {Heinz}}]{mak_control_2012}%
  \BibitemOpen
  \bibfield  {author} {\bibinfo {author} {\bibfnamefont {K.~F.}\ \bibnamefont {Mak}}, \bibinfo {author} {\bibfnamefont {K.}~\bibnamefont {He}}, \bibinfo {author} {\bibfnamefont {J.}~\bibnamefont {Shan}}, \ and\ \bibinfo {author} {\bibfnamefont {T.~F.}\ \bibnamefont {Heinz}},\ }\href {\doibase 10.1038/nnano.2012.96} {\bibfield  {journal} {\bibinfo  {journal} {Nature Nanotechnology}\ }\textbf {\bibinfo {volume} {7}},\ \bibinfo {pages} {494} (\bibinfo {year} {2012})}\BibitemShut {NoStop}%
\bibitem [{\citenamefont {Riley}\ \emph {et~al.}(2014)\citenamefont {Riley}, \citenamefont {Mazzola}, \citenamefont {Dendzik}, \citenamefont {Michiardi}, \citenamefont {Takayama}, \citenamefont {Bawden}, \citenamefont {Graner{\o}d}, \citenamefont {Leandersson}, \citenamefont {Balasubramanian}, \citenamefont {Hoesch}, \citenamefont {Kim}, \citenamefont {Takagi}, \citenamefont {Meevasana}, \citenamefont {Hofmann}, \citenamefont {Bahramy}, \citenamefont {Wells},\ and\ \citenamefont {King}}]{riley_direct_2014}%
  \BibitemOpen
  \bibfield  {author} {\bibinfo {author} {\bibfnamefont {J.~M.}\ \bibnamefont {Riley}}, \bibinfo {author} {\bibfnamefont {F.}~\bibnamefont {Mazzola}}, \bibinfo {author} {\bibfnamefont {M.}~\bibnamefont {Dendzik}}, \bibinfo {author} {\bibfnamefont {M.}~\bibnamefont {Michiardi}}, \bibinfo {author} {\bibfnamefont {T.}~\bibnamefont {Takayama}}, \bibinfo {author} {\bibfnamefont {L.}~\bibnamefont {Bawden}}, \bibinfo {author} {\bibfnamefont {C.}~\bibnamefont {Graner{\o}d}}, \bibinfo {author} {\bibfnamefont {M.}~\bibnamefont {Leandersson}}, \bibinfo {author} {\bibfnamefont {T.}~\bibnamefont {Balasubramanian}}, \bibinfo {author} {\bibfnamefont {M.}~\bibnamefont {Hoesch}}, \bibinfo {author} {\bibfnamefont {T.~K.}\ \bibnamefont {Kim}}, \bibinfo {author} {\bibfnamefont {H.}~\bibnamefont {Takagi}}, \bibinfo {author} {\bibfnamefont {W.}~\bibnamefont {Meevasana}}, \bibinfo {author} {\bibfnamefont {P.}~\bibnamefont {Hofmann}}, \bibinfo {author} {\bibfnamefont {M.~S.}\ \bibnamefont {Bahramy}}, \bibinfo {author} {\bibfnamefont
  {J.~W.}\ \bibnamefont {Wells}}, \ and\ \bibinfo {author} {\bibfnamefont {P.~D.~C.}\ \bibnamefont {King}},\ }\href {\doibase 10.1038/nphys3105} {\bibfield  {journal} {\bibinfo  {journal} {Nature Physics}\ }\textbf {\bibinfo {volume} {10}},\ \bibinfo {pages} {835} (\bibinfo {year} {2014})}\BibitemShut {NoStop}%
\bibitem [{\citenamefont {Yuan}\ \emph {et~al.}(2013)\citenamefont {Yuan}, \citenamefont {Bahramy}, \citenamefont {Morimoto}, \citenamefont {Wu}, \citenamefont {Nomura}, \citenamefont {Yang}, \citenamefont {Shimotani}, \citenamefont {Suzuki}, \citenamefont {Toh}, \citenamefont {Kloc}, \citenamefont {Xu}, \citenamefont {Arita}, \citenamefont {Nagaosa},\ and\ \citenamefont {Iwasa}}]{yuan_zeeman_2013}%
  \BibitemOpen
  \bibfield  {author} {\bibinfo {author} {\bibfnamefont {H.}~\bibnamefont {Yuan}}, \bibinfo {author} {\bibfnamefont {M.~S.}\ \bibnamefont {Bahramy}}, \bibinfo {author} {\bibfnamefont {K.}~\bibnamefont {Morimoto}}, \bibinfo {author} {\bibfnamefont {S.}~\bibnamefont {Wu}}, \bibinfo {author} {\bibfnamefont {K.}~\bibnamefont {Nomura}}, \bibinfo {author} {\bibfnamefont {B.-J.}\ \bibnamefont {Yang}}, \bibinfo {author} {\bibfnamefont {H.}~\bibnamefont {Shimotani}}, \bibinfo {author} {\bibfnamefont {R.}~\bibnamefont {Suzuki}}, \bibinfo {author} {\bibfnamefont {M.}~\bibnamefont {Toh}}, \bibinfo {author} {\bibfnamefont {C.}~\bibnamefont {Kloc}}, \bibinfo {author} {\bibfnamefont {X.}~\bibnamefont {Xu}}, \bibinfo {author} {\bibfnamefont {R.}~\bibnamefont {Arita}}, \bibinfo {author} {\bibfnamefont {N.}~\bibnamefont {Nagaosa}}, \ and\ \bibinfo {author} {\bibfnamefont {Y.}~\bibnamefont {Iwasa}},\ }\href {\doibase 10.1038/nphys2691} {\bibfield  {journal} {\bibinfo  {journal} {Nature Physics}\ }\textbf {\bibinfo {volume}
  {9}},\ \bibinfo {pages} {563} (\bibinfo {year} {2013})}\BibitemShut {NoStop}%
\bibitem [{\citenamefont {Tonndorf}\ \emph {et~al.}(2013)\citenamefont {Tonndorf}, \citenamefont {Schmidt}, \citenamefont {B{\"o}ttger}, \citenamefont {Zhang}, \citenamefont {B{\"o}rner}, \citenamefont {Liebig}, \citenamefont {Albrecht}, \citenamefont {Kloc}, \citenamefont {Gordan}, \citenamefont {Zahn}, \citenamefont {Michaelis~de Vasconcellos},\ and\ \citenamefont {Bratschitsch}}]{tonndorf_photoluminesence_2013}%
  \BibitemOpen
  \bibfield  {author} {\bibinfo {author} {\bibfnamefont {P.}~\bibnamefont {Tonndorf}}, \bibinfo {author} {\bibfnamefont {R.}~\bibnamefont {Schmidt}}, \bibinfo {author} {\bibfnamefont {P.}~\bibnamefont {B{\"o}ttger}}, \bibinfo {author} {\bibfnamefont {X.}~\bibnamefont {Zhang}}, \bibinfo {author} {\bibfnamefont {J.}~\bibnamefont {B{\"o}rner}}, \bibinfo {author} {\bibfnamefont {A.}~\bibnamefont {Liebig}}, \bibinfo {author} {\bibfnamefont {M.}~\bibnamefont {Albrecht}}, \bibinfo {author} {\bibfnamefont {C.}~\bibnamefont {Kloc}}, \bibinfo {author} {\bibfnamefont {O.}~\bibnamefont {Gordan}}, \bibinfo {author} {\bibfnamefont {D.~R.~T.}\ \bibnamefont {Zahn}}, \bibinfo {author} {\bibfnamefont {S.}~\bibnamefont {Michaelis~de Vasconcellos}}, \ and\ \bibinfo {author} {\bibfnamefont {R.}~\bibnamefont {Bratschitsch}},\ }\href {\doibase 10.1364/OE.21.004908} {\bibfield  {journal} {\bibinfo  {journal} {Optics Express}\ }\textbf {\bibinfo {volume} {21}},\ \bibinfo {pages} {4908} (\bibinfo {year} {2013})}\BibitemShut {NoStop}%
\bibitem [{\citenamefont {Chhowalla}\ \emph {et~al.}(2013)\citenamefont {Chhowalla}, \citenamefont {Shin}, \citenamefont {Eda}, \citenamefont {Li}, \citenamefont {Loh},\ and\ \citenamefont {Zhang}}]{chhowalla_chemistry_2013}%
  \BibitemOpen
  \bibfield  {author} {\bibinfo {author} {\bibfnamefont {M.}~\bibnamefont {Chhowalla}}, \bibinfo {author} {\bibfnamefont {H.~S.}\ \bibnamefont {Shin}}, \bibinfo {author} {\bibfnamefont {G.}~\bibnamefont {Eda}}, \bibinfo {author} {\bibfnamefont {L.-J.}\ \bibnamefont {Li}}, \bibinfo {author} {\bibfnamefont {K.~P.}\ \bibnamefont {Loh}}, \ and\ \bibinfo {author} {\bibfnamefont {H.}~\bibnamefont {Zhang}},\ }\href {\doibase 10.1038/nchem.1589} {\bibfield  {journal} {\bibinfo  {journal} {Nature Chemistry}\ }\textbf {\bibinfo {volume} {5}},\ \bibinfo {pages} {263} (\bibinfo {year} {2013})}\BibitemShut {NoStop}%
\bibitem [{\citenamefont {Borisenko}\ \emph {et~al.}(2009)\citenamefont {Borisenko}, \citenamefont {Kordyuk}, \citenamefont {Zabolotnyy}, \citenamefont {Inosov}, \citenamefont {Evtushinsky}, \citenamefont {B\"uchner}, \citenamefont {Yaresko}, \citenamefont {Varykhalov}, \citenamefont {Follath}, \citenamefont {Eberhardt}, \citenamefont {Patthey},\ and\ \citenamefont {Berger}}]{borisenko_two_2009}%
  \BibitemOpen
  \bibfield  {author} {\bibinfo {author} {\bibfnamefont {S.~V.}\ \bibnamefont {Borisenko}}, \bibinfo {author} {\bibfnamefont {A.~A.}\ \bibnamefont {Kordyuk}}, \bibinfo {author} {\bibfnamefont {V.~B.}\ \bibnamefont {Zabolotnyy}}, \bibinfo {author} {\bibfnamefont {D.~S.}\ \bibnamefont {Inosov}}, \bibinfo {author} {\bibfnamefont {D.}~\bibnamefont {Evtushinsky}}, \bibinfo {author} {\bibfnamefont {B.}~\bibnamefont {B\"uchner}}, \bibinfo {author} {\bibfnamefont {A.~N.}\ \bibnamefont {Yaresko}}, \bibinfo {author} {\bibfnamefont {A.}~\bibnamefont {Varykhalov}}, \bibinfo {author} {\bibfnamefont {R.}~\bibnamefont {Follath}}, \bibinfo {author} {\bibfnamefont {W.}~\bibnamefont {Eberhardt}}, \bibinfo {author} {\bibfnamefont {L.}~\bibnamefont {Patthey}}, \ and\ \bibinfo {author} {\bibfnamefont {H.}~\bibnamefont {Berger}},\ }\href {\doibase 10.1103/PhysRevLett.102.166402} {\bibfield  {journal} {\bibinfo  {journal} {Phys. Rev. Lett.}\ }\textbf {\bibinfo {volume} {102}},\ \bibinfo {pages} {166402} (\bibinfo {year}
  {2009})}\BibitemShut {NoStop}%
\bibitem [{\citenamefont {Bawden}\ \emph {et~al.}(2016)\citenamefont {Bawden}, \citenamefont {Cooil}, \citenamefont {Mazzola}, \citenamefont {Riley}, \citenamefont {Collins-McIntyre}, \citenamefont {Sunko}, \citenamefont {Hunvik}, \citenamefont {Leandersson}, \citenamefont {Polley}, \citenamefont {Balasubramanian}, \citenamefont {Kim}, \citenamefont {Hoesch}, \citenamefont {Wells}, \citenamefont {Balakrishnan}, \citenamefont {Bahramy},\ and\ \citenamefont {King}}]{bawden_spin_2015}%
  \BibitemOpen
  \bibfield  {author} {\bibinfo {author} {\bibfnamefont {L.}~\bibnamefont {Bawden}}, \bibinfo {author} {\bibfnamefont {S.~P.}\ \bibnamefont {Cooil}}, \bibinfo {author} {\bibfnamefont {F.}~\bibnamefont {Mazzola}}, \bibinfo {author} {\bibfnamefont {J.~M.}\ \bibnamefont {Riley}}, \bibinfo {author} {\bibfnamefont {L.~J.}\ \bibnamefont {Collins-McIntyre}}, \bibinfo {author} {\bibfnamefont {V.}~\bibnamefont {Sunko}}, \bibinfo {author} {\bibfnamefont {K.~W.~B.}\ \bibnamefont {Hunvik}}, \bibinfo {author} {\bibfnamefont {M.}~\bibnamefont {Leandersson}}, \bibinfo {author} {\bibfnamefont {C.~M.}\ \bibnamefont {Polley}}, \bibinfo {author} {\bibfnamefont {T.}~\bibnamefont {Balasubramanian}}, \bibinfo {author} {\bibfnamefont {T.~K.}\ \bibnamefont {Kim}}, \bibinfo {author} {\bibfnamefont {M.}~\bibnamefont {Hoesch}}, \bibinfo {author} {\bibfnamefont {J.~W.}\ \bibnamefont {Wells}}, \bibinfo {author} {\bibfnamefont {G.}~\bibnamefont {Balakrishnan}}, \bibinfo {author} {\bibfnamefont {M.~S.}\ \bibnamefont {Bahramy}}, \ and\
  \bibinfo {author} {\bibfnamefont {P.~D.~C.}\ \bibnamefont {King}},\ }\href {\doibase 10.1038/ncomms11711} {\bibfield  {journal} {\bibinfo  {journal} {Nature Communications}\ }\textbf {\bibinfo {volume} {7}},\ \bibinfo {pages} {11711} (\bibinfo {year} {2016})}\BibitemShut {NoStop}%
\bibitem [{\citenamefont {Riley}\ \emph {et~al.}(2015)\citenamefont {Riley}, \citenamefont {Meevasana}, \citenamefont {Bawden}, \citenamefont {Asakawa}, \citenamefont {Takayama}, \citenamefont {Eknapakul}, \citenamefont {Kim}, \citenamefont {Hoesch}, \citenamefont {Mo}, \citenamefont {Takagi}, \citenamefont {Sasagawa}, \citenamefont {Bahramy},\ and\ \citenamefont {King}}]{riley_negative_2015}%
  \BibitemOpen
  \bibfield  {author} {\bibinfo {author} {\bibfnamefont {J.~M.}\ \bibnamefont {Riley}}, \bibinfo {author} {\bibfnamefont {W.}~\bibnamefont {Meevasana}}, \bibinfo {author} {\bibfnamefont {L.}~\bibnamefont {Bawden}}, \bibinfo {author} {\bibfnamefont {M.}~\bibnamefont {Asakawa}}, \bibinfo {author} {\bibfnamefont {T.}~\bibnamefont {Takayama}}, \bibinfo {author} {\bibfnamefont {T.}~\bibnamefont {Eknapakul}}, \bibinfo {author} {\bibfnamefont {T.~K.}\ \bibnamefont {Kim}}, \bibinfo {author} {\bibfnamefont {M.}~\bibnamefont {Hoesch}}, \bibinfo {author} {\bibfnamefont {S.~K.}\ \bibnamefont {Mo}}, \bibinfo {author} {\bibfnamefont {H.}~\bibnamefont {Takagi}}, \bibinfo {author} {\bibfnamefont {T.}~\bibnamefont {Sasagawa}}, \bibinfo {author} {\bibfnamefont {M.~S.}\ \bibnamefont {Bahramy}}, \ and\ \bibinfo {author} {\bibfnamefont {P.~D.~C.}\ \bibnamefont {King}},\ }\href {\doibase 10.1038/nnano.2015.217} {\bibfield  {journal} {\bibinfo  {journal} {Nature Nanotechnology}\ }\textbf {\bibinfo {volume} {10}},\ \bibinfo {pages}
  {1043} (\bibinfo {year} {2015})}\BibitemShut {NoStop}%
\bibitem [{\citenamefont {Nguyen}\ \emph {et~al.}(2019)\citenamefont {Nguyen}, \citenamefont {Teutsch}, \citenamefont {Wilson}, \citenamefont {Kahn}, \citenamefont {Xia}, \citenamefont {Graham}, \citenamefont {Kandyba}, \citenamefont {Giampietri}, \citenamefont {Barinov}, \citenamefont {Constantinescu}, \citenamefont {Yeung}, \citenamefont {Hine}, \citenamefont {Xu}, \citenamefont {Cobden},\ and\ \citenamefont {Wilson}}]{nguyen_visualizing_2019}%
  \BibitemOpen
  \bibfield  {author} {\bibinfo {author} {\bibfnamefont {P.~V.}\ \bibnamefont {Nguyen}}, \bibinfo {author} {\bibfnamefont {N.~C.}\ \bibnamefont {Teutsch}}, \bibinfo {author} {\bibfnamefont {N.~P.}\ \bibnamefont {Wilson}}, \bibinfo {author} {\bibfnamefont {J.}~\bibnamefont {Kahn}}, \bibinfo {author} {\bibfnamefont {X.}~\bibnamefont {Xia}}, \bibinfo {author} {\bibfnamefont {A.~J.}\ \bibnamefont {Graham}}, \bibinfo {author} {\bibfnamefont {V.}~\bibnamefont {Kandyba}}, \bibinfo {author} {\bibfnamefont {A.}~\bibnamefont {Giampietri}}, \bibinfo {author} {\bibfnamefont {A.}~\bibnamefont {Barinov}}, \bibinfo {author} {\bibfnamefont {G.~C.}\ \bibnamefont {Constantinescu}}, \bibinfo {author} {\bibfnamefont {N.}~\bibnamefont {Yeung}}, \bibinfo {author} {\bibfnamefont {N.~D.~M.}\ \bibnamefont {Hine}}, \bibinfo {author} {\bibfnamefont {X.}~\bibnamefont {Xu}}, \bibinfo {author} {\bibfnamefont {D.~H.}\ \bibnamefont {Cobden}}, \ and\ \bibinfo {author} {\bibfnamefont {N.~R.}\ \bibnamefont {Wilson}},\ }\href {\doibase
  10.1038/s41586-019-1402-1} {\bibfield  {journal} {\bibinfo  {journal} {Nature}\ }\textbf {\bibinfo {volume} {572}},\ \bibinfo {pages} {220} (\bibinfo {year} {2019})}\BibitemShut {NoStop}%
\bibitem [{\citenamefont {Damascelli}(2004)}]{Dam2004}%
  \BibitemOpen
  \bibfield  {author} {\bibinfo {author} {\bibfnamefont {A.}~\bibnamefont {Damascelli}},\ }\href@noop {} {\bibfield  {journal} {\bibinfo  {journal} {Physica Scripta}\ }\textbf {\bibinfo {volume} {2004}},\ \bibinfo {pages} {T109 61} (\bibinfo {year} {2004})}\BibitemShut {NoStop}%
\bibitem [{\citenamefont {Devakul}\ \emph {et~al.}(2021)\citenamefont {Devakul}, \citenamefont {Cr{\'e}pel}, \citenamefont {Zhang},\ and\ \citenamefont {Fu}}]{devakul_magic_2021}%
  \BibitemOpen
  \bibfield  {author} {\bibinfo {author} {\bibfnamefont {T.}~\bibnamefont {Devakul}}, \bibinfo {author} {\bibfnamefont {V.}~\bibnamefont {Cr{\'e}pel}}, \bibinfo {author} {\bibfnamefont {Y.}~\bibnamefont {Zhang}}, \ and\ \bibinfo {author} {\bibfnamefont {L.}~\bibnamefont {Fu}},\ }\href {\doibase 10.1038/s41467-021-27042-9} {\bibfield  {journal} {\bibinfo  {journal} {Nature Communications}\ }\textbf {\bibinfo {volume} {12}},\ \bibinfo {pages} {6730} (\bibinfo {year} {2021})}\BibitemShut {NoStop}%
\bibitem [{\citenamefont {Guo}\ \emph {et~al.}(2023)\citenamefont {Guo}, \citenamefont {Pack}, \citenamefont {Swann}, \citenamefont {Holtzman}, \citenamefont {Cothrine}, \citenamefont {Watanabe}, \citenamefont {Taniguchi}, \citenamefont {Mandrus}, \citenamefont {Barmak}, \citenamefont {Hone}, \citenamefont {Millis}, \citenamefont {Pasupathy},\ and\ \citenamefont {Dean}}]{guo_superconductivity_2024}%
  \BibitemOpen
  \bibfield  {author} {\bibinfo {author} {\bibfnamefont {Y.}~\bibnamefont {Guo}}, \bibinfo {author} {\bibfnamefont {J.}~\bibnamefont {Pack}}, \bibinfo {author} {\bibfnamefont {J.}~\bibnamefont {Swann}}, \bibinfo {author} {\bibfnamefont {L.}~\bibnamefont {Holtzman}}, \bibinfo {author} {\bibfnamefont {M.}~\bibnamefont {Cothrine}}, \bibinfo {author} {\bibfnamefont {K.}~\bibnamefont {Watanabe}}, \bibinfo {author} {\bibfnamefont {T.}~\bibnamefont {Taniguchi}}, \bibinfo {author} {\bibfnamefont {D.}~\bibnamefont {Mandrus}}, \bibinfo {author} {\bibfnamefont {K.}~\bibnamefont {Barmak}}, \bibinfo {author} {\bibfnamefont {J.}~\bibnamefont {Hone}}, \bibinfo {author} {\bibfnamefont {A.~J.}\ \bibnamefont {Millis}}, \bibinfo {author} {\bibfnamefont {A.~N.}\ \bibnamefont {Pasupathy}}, \ and\ \bibinfo {author} {\bibfnamefont {C.~R.}\ \bibnamefont {Dean}},\ }\href@noop {} {\bibfield  {journal} {\bibinfo  {journal} {arXiv:2310.11317}\ } (\bibinfo {year} {2023})}\BibitemShut {NoStop}%
\bibitem [{\citenamefont {Bahramy}\ \emph {et~al.}(2018)\citenamefont {Bahramy}, \citenamefont {Clark}, \citenamefont {Yang}, \citenamefont {Feng}, \citenamefont {Bawden}, \citenamefont {Riley}, \citenamefont {I.}, \citenamefont {Mazzola}, \citenamefont {Sunko}, \citenamefont {Biswas}, \citenamefont {Cooil}, \citenamefont {Jorge}, \citenamefont {Wells}, \citenamefont {Leandersson}, \citenamefont {Balasubramanian}, \citenamefont {Fujii}, \citenamefont {Vobornik}, \citenamefont {Rault}, \citenamefont {Kim}, \citenamefont {Hoesch}, \citenamefont {Okawa}, \citenamefont {Asakawa}, \citenamefont {Sasagawa}, \citenamefont {Eknapakul}, \citenamefont {Meevasana},\ and\ \citenamefont {King}}]{bahramy_ubiquitous_2018}%
  \BibitemOpen
  \bibfield  {author} {\bibinfo {author} {\bibfnamefont {M.~S.}\ \bibnamefont {Bahramy}}, \bibinfo {author} {\bibfnamefont {O.~J.}\ \bibnamefont {Clark}}, \bibinfo {author} {\bibfnamefont {B.-J.}\ \bibnamefont {Yang}}, \bibinfo {author} {\bibfnamefont {J.}~\bibnamefont {Feng}}, \bibinfo {author} {\bibfnamefont {L.}~\bibnamefont {Bawden}}, \bibinfo {author} {\bibfnamefont {J.~M.}\ \bibnamefont {Riley}}, \bibinfo {author} {\bibfnamefont {M.}~\bibnamefont {I.}}, \bibinfo {author} {\bibfnamefont {F.}~\bibnamefont {Mazzola}}, \bibinfo {author} {\bibfnamefont {V.}~\bibnamefont {Sunko}}, \bibinfo {author} {\bibfnamefont {D.}~\bibnamefont {Biswas}}, \bibinfo {author} {\bibfnamefont {S.~P.}\ \bibnamefont {Cooil}}, \bibinfo {author} {\bibfnamefont {M.}~\bibnamefont {Jorge}}, \bibinfo {author} {\bibfnamefont {J.~W.}\ \bibnamefont {Wells}}, \bibinfo {author} {\bibfnamefont {M.}~\bibnamefont {Leandersson}}, \bibinfo {author} {\bibfnamefont {T.}~\bibnamefont {Balasubramanian}}, \bibinfo {author} {\bibfnamefont
  {J.}~\bibnamefont {Fujii}}, \bibinfo {author} {\bibfnamefont {I.}~\bibnamefont {Vobornik}}, \bibinfo {author} {\bibfnamefont {J.}~\bibnamefont {Rault}}, \bibinfo {author} {\bibfnamefont {T.~K.}\ \bibnamefont {Kim}}, \bibinfo {author} {\bibfnamefont {M.}~\bibnamefont {Hoesch}}, \bibinfo {author} {\bibfnamefont {K.}~\bibnamefont {Okawa}}, \bibinfo {author} {\bibfnamefont {M.}~\bibnamefont {Asakawa}}, \bibinfo {author} {\bibfnamefont {T.}~\bibnamefont {Sasagawa}}, \bibinfo {author} {\bibfnamefont {T.}~\bibnamefont {Eknapakul}}, \bibinfo {author} {\bibfnamefont {W.}~\bibnamefont {Meevasana}}, \ and\ \bibinfo {author} {\bibfnamefont {P.~D.~C.}\ \bibnamefont {King}},\ }\href@noop {} {\bibfield  {journal} {\bibinfo  {journal} {Nature Materials}\ }\textbf {\bibinfo {volume} {17}} (\bibinfo {year} {2018})}\BibitemShut {NoStop}%
\bibitem [{\citenamefont {Zhai}\ \emph {et~al.}()\citenamefont {Zhai}, \citenamefont {Baniya}, \citenamefont {Zhang}, \citenamefont {Li}, \citenamefont {Haney}, \citenamefont {Sheng}, \citenamefont {Ehrenfreund},\ and\ \citenamefont {Vardeny}}]{zhai_transient_2017}%
  \BibitemOpen
  \bibfield  {author} {\bibinfo {author} {\bibfnamefont {Y.}~\bibnamefont {Zhai}}, \bibinfo {author} {\bibfnamefont {S.}~\bibnamefont {Baniya}}, \bibinfo {author} {\bibfnamefont {C.}~\bibnamefont {Zhang}}, \bibinfo {author} {\bibfnamefont {J.}~\bibnamefont {Li}}, \bibinfo {author} {\bibfnamefont {P.}~\bibnamefont {Haney}}, \bibinfo {author} {\bibfnamefont {C.-X.}\ \bibnamefont {Sheng}}, \bibinfo {author} {\bibfnamefont {E.}~\bibnamefont {Ehrenfreund}}, \ and\ \bibinfo {author} {\bibfnamefont {Z.~V.}\ \bibnamefont {Vardeny}},\ }\href {\doibase 10.1126/sciadv.1700704} {\bibfield  {journal} {\bibinfo  {journal} {Science Advances}\ }\textbf {\bibinfo {volume} {3}},\ \bibinfo {pages} {e1700704}}\BibitemShut {NoStop}%
\bibitem [{\citenamefont {Sunko}\ \emph {et~al.}(2017)\citenamefont {Sunko}, \citenamefont {Rosner}, \citenamefont {Kushwaha}, \citenamefont {Khim}, \citenamefont {Mazzola}, \citenamefont {Bawden}, \citenamefont {Clark}, \citenamefont {Riley}, \citenamefont {Kasinathan}, \citenamefont {Haverkort}, \citenamefont {Kim}, \citenamefont {Hoesch}, \citenamefont {Fujii}, \citenamefont {Vobornik}, \citenamefont {Mackenzie},\ and\ \citenamefont {King}}]{sunko_maximal_2017}%
  \BibitemOpen
  \bibfield  {author} {\bibinfo {author} {\bibfnamefont {V.}~\bibnamefont {Sunko}}, \bibinfo {author} {\bibfnamefont {H.}~\bibnamefont {Rosner}}, \bibinfo {author} {\bibfnamefont {P.}~\bibnamefont {Kushwaha}}, \bibinfo {author} {\bibfnamefont {S.}~\bibnamefont {Khim}}, \bibinfo {author} {\bibfnamefont {F.}~\bibnamefont {Mazzola}}, \bibinfo {author} {\bibfnamefont {L.}~\bibnamefont {Bawden}}, \bibinfo {author} {\bibfnamefont {O.~J.}\ \bibnamefont {Clark}}, \bibinfo {author} {\bibfnamefont {J.~M.}\ \bibnamefont {Riley}}, \bibinfo {author} {\bibfnamefont {D.}~\bibnamefont {Kasinathan}}, \bibinfo {author} {\bibfnamefont {M.~W.}\ \bibnamefont {Haverkort}}, \bibinfo {author} {\bibfnamefont {T.~K.}\ \bibnamefont {Kim}}, \bibinfo {author} {\bibfnamefont {M.}~\bibnamefont {Hoesch}}, \bibinfo {author} {\bibfnamefont {J.}~\bibnamefont {Fujii}}, \bibinfo {author} {\bibfnamefont {I.}~\bibnamefont {Vobornik}}, \bibinfo {author} {\bibfnamefont {A.~P.}\ \bibnamefont {Mackenzie}}, \ and\ \bibinfo {author} {\bibfnamefont
  {P.~D.~C.}\ \bibnamefont {King}},\ }\href {\doibase 10.1038/nature23898} {\bibfield  {journal} {\bibinfo  {journal} {Nature}\ }\textbf {\bibinfo {volume} {549}},\ \bibinfo {pages} {492} (\bibinfo {year} {2017})}\BibitemShut {NoStop}%
\bibitem [{\citenamefont {Ishizaka}\ \emph {et~al.}(2011)\citenamefont {Ishizaka}, \citenamefont {Bahramy}, \citenamefont {Murakawa}, \citenamefont {Sakano}, \citenamefont {Shimojima}, \citenamefont {Sonobe}, \citenamefont {Koizumi}, \citenamefont {Shin}, \citenamefont {Miyahara}, \citenamefont {Kimura}, \citenamefont {Miyamoto}, \citenamefont {Okuda}, \citenamefont {Namatame}, \citenamefont {Taniguchi}, \citenamefont {Arita}, \citenamefont {Nagaosa}, \citenamefont {Kobayashi}, \citenamefont {Murakami}, \citenamefont {Kumai}, \citenamefont {Kaneko}, \citenamefont {Onose},\ and\ \citenamefont {Tokura}}]{ishizaka_giant_2011}%
  \BibitemOpen
  \bibfield  {author} {\bibinfo {author} {\bibfnamefont {K.}~\bibnamefont {Ishizaka}}, \bibinfo {author} {\bibfnamefont {M.~S.}\ \bibnamefont {Bahramy}}, \bibinfo {author} {\bibfnamefont {H.}~\bibnamefont {Murakawa}}, \bibinfo {author} {\bibfnamefont {M.}~\bibnamefont {Sakano}}, \bibinfo {author} {\bibfnamefont {T.}~\bibnamefont {Shimojima}}, \bibinfo {author} {\bibfnamefont {T.}~\bibnamefont {Sonobe}}, \bibinfo {author} {\bibfnamefont {K.}~\bibnamefont {Koizumi}}, \bibinfo {author} {\bibfnamefont {S.}~\bibnamefont {Shin}}, \bibinfo {author} {\bibfnamefont {H.}~\bibnamefont {Miyahara}}, \bibinfo {author} {\bibfnamefont {A.}~\bibnamefont {Kimura}}, \bibinfo {author} {\bibfnamefont {K.}~\bibnamefont {Miyamoto}}, \bibinfo {author} {\bibfnamefont {T.}~\bibnamefont {Okuda}}, \bibinfo {author} {\bibfnamefont {H.}~\bibnamefont {Namatame}}, \bibinfo {author} {\bibfnamefont {M.}~\bibnamefont {Taniguchi}}, \bibinfo {author} {\bibfnamefont {R.}~\bibnamefont {Arita}}, \bibinfo {author} {\bibfnamefont {N.}~\bibnamefont
  {Nagaosa}}, \bibinfo {author} {\bibfnamefont {K.}~\bibnamefont {Kobayashi}}, \bibinfo {author} {\bibfnamefont {Y.}~\bibnamefont {Murakami}}, \bibinfo {author} {\bibfnamefont {R.}~\bibnamefont {Kumai}}, \bibinfo {author} {\bibfnamefont {Y.}~\bibnamefont {Kaneko}}, \bibinfo {author} {\bibfnamefont {Y.}~\bibnamefont {Onose}}, \ and\ \bibinfo {author} {\bibfnamefont {Y.}~\bibnamefont {Tokura}},\ }\href {\doibase 10.1038/nmat3051} {\bibfield  {journal} {\bibinfo  {journal} {Nature Materials}\ }\textbf {\bibinfo {volume} {10}},\ \bibinfo {pages} {521} (\bibinfo {year} {2011})}\BibitemShut {NoStop}%
\bibitem [{\citenamefont {Bian}\ \emph {et~al.}(2013)\citenamefont {Bian}, \citenamefont {Wang}, \citenamefont {Miller},\ and\ \citenamefont {Chiang}}]{bian_origin_2013}%
  \BibitemOpen
  \bibfield  {author} {\bibinfo {author} {\bibfnamefont {G.}~\bibnamefont {Bian}}, \bibinfo {author} {\bibfnamefont {X.}~\bibnamefont {Wang}}, \bibinfo {author} {\bibfnamefont {T.}~\bibnamefont {Miller}}, \ and\ \bibinfo {author} {\bibfnamefont {T.-C.}\ \bibnamefont {Chiang}},\ }\href {\doibase 10.1103/PhysRevB.88.085427} {\bibfield  {journal} {\bibinfo  {journal} {Phys. Rev. B}\ }\textbf {\bibinfo {volume} {88}},\ \bibinfo {pages} {085427} (\bibinfo {year} {2013})}\BibitemShut {NoStop}%
\bibitem [{\citenamefont {Bianchi}\ \emph {et~al.}(2012)\citenamefont {Bianchi}, \citenamefont {Hatch}, \citenamefont {Li}, \citenamefont {Hofmann}, \citenamefont {Song}, \citenamefont {Mi}, \citenamefont {Iversen}, \citenamefont {Abd El-Fattah}, \citenamefont {L{\"o}ptien}, \citenamefont {Zhou}, \citenamefont {Khajetoorians}, \citenamefont {Wiebe}, \citenamefont {Wiesendanger},\ and\ \citenamefont {Wells}}]{bianchi_robust_2012}%
  \BibitemOpen
  \bibfield  {author} {\bibinfo {author} {\bibfnamefont {M.}~\bibnamefont {Bianchi}}, \bibinfo {author} {\bibfnamefont {R.~C.}\ \bibnamefont {Hatch}}, \bibinfo {author} {\bibfnamefont {Z.}~\bibnamefont {Li}}, \bibinfo {author} {\bibfnamefont {P.}~\bibnamefont {Hofmann}}, \bibinfo {author} {\bibfnamefont {F.}~\bibnamefont {Song}}, \bibinfo {author} {\bibfnamefont {J.}~\bibnamefont {Mi}}, \bibinfo {author} {\bibfnamefont {B.~B.}\ \bibnamefont {Iversen}}, \bibinfo {author} {\bibfnamefont {Z.~M.}\ \bibnamefont {Abd El-Fattah}}, \bibinfo {author} {\bibfnamefont {P.}~\bibnamefont {L{\"o}ptien}}, \bibinfo {author} {\bibfnamefont {L.}~\bibnamefont {Zhou}}, \bibinfo {author} {\bibfnamefont {A.~A.}\ \bibnamefont {Khajetoorians}}, \bibinfo {author} {\bibfnamefont {J.}~\bibnamefont {Wiebe}}, \bibinfo {author} {\bibfnamefont {R.}~\bibnamefont {Wiesendanger}}, \ and\ \bibinfo {author} {\bibfnamefont {J.~W.}\ \bibnamefont {Wells}},\ }\href {\doibase 10.1021/nn3021822} {\bibfield  {journal} {\bibinfo  {journal} {ACS Nano}\
  }\textbf {\bibinfo {volume} {6}},\ \bibinfo {pages} {7009} (\bibinfo {year} {2012})}\BibitemShut {NoStop}%
\bibitem [{\citenamefont {Feng}\ \emph {et~al.}(2025)\citenamefont {Feng}, \citenamefont {Zhang}, \citenamefont {Li}, \citenamefont {Li}, \citenamefont {Bao}, \citenamefont {Zhang}, \citenamefont {Chen}, \citenamefont {Tang}, \citenamefont {Yaegashi}, \citenamefont {Sugawara}, \citenamefont {Sato}, \citenamefont {Duan}, \citenamefont {Yu},\ and\ \citenamefont {Zhou}}]{feng_giant_2025}%
  \BibitemOpen
  \bibfield  {author} {\bibinfo {author} {\bibfnamefont {R.}~\bibnamefont {Feng}}, \bibinfo {author} {\bibfnamefont {Y.}~\bibnamefont {Zhang}}, \bibinfo {author} {\bibfnamefont {J.}~\bibnamefont {Li}}, \bibinfo {author} {\bibfnamefont {Q.}~\bibnamefont {Li}}, \bibinfo {author} {\bibfnamefont {C.}~\bibnamefont {Bao}}, \bibinfo {author} {\bibfnamefont {H.}~\bibnamefont {Zhang}}, \bibinfo {author} {\bibfnamefont {W.}~\bibnamefont {Chen}}, \bibinfo {author} {\bibfnamefont {X.}~\bibnamefont {Tang}}, \bibinfo {author} {\bibfnamefont {K.}~\bibnamefont {Yaegashi}}, \bibinfo {author} {\bibfnamefont {K.}~\bibnamefont {Sugawara}}, \bibinfo {author} {\bibfnamefont {T.}~\bibnamefont {Sato}}, \bibinfo {author} {\bibfnamefont {W.}~\bibnamefont {Duan}}, \bibinfo {author} {\bibfnamefont {P.}~\bibnamefont {Yu}}, \ and\ \bibinfo {author} {\bibfnamefont {S.}~\bibnamefont {Zhou}},\ }\href {\doibase 10.1038/s41467-025-57835-1} {\bibfield  {journal} {\bibinfo  {journal} {Nature Communications}\ }\textbf {\bibinfo {volume} {16}},\
  \bibinfo {pages} {2667} (\bibinfo {year} {2025})}\BibitemShut {NoStop}%
\bibitem [{\citenamefont {Picozzi}(2014)}]{picozzi_ferroelectric_2014}%
  \BibitemOpen
  \bibfield  {author} {\bibinfo {author} {\bibfnamefont {S.}~\bibnamefont {Picozzi}},\ }\href {https://www.frontiersin.org/journals/physics/articles/10.3389/fphy.2014.00010} {\bibfield  {journal} {\bibinfo  {journal} {Frontiers in Physics}\ }\textbf {\bibinfo {volume} {Volume 2 - 2014}} (\bibinfo {year} {2014})}\BibitemShut {NoStop}%
\bibitem [{\citenamefont {Hasan}\ and\ \citenamefont {Kane}(2010)}]{hasan_colloquium_2010}%
  \BibitemOpen
  \bibfield  {author} {\bibinfo {author} {\bibfnamefont {M.~Z.}\ \bibnamefont {Hasan}}\ and\ \bibinfo {author} {\bibfnamefont {C.~L.}\ \bibnamefont {Kane}},\ }\href {\doibase 10.1103/RevModPhys.82.3045} {\bibfield  {journal} {\bibinfo  {journal} {Rev. Mod. Phys.}\ }\textbf {\bibinfo {volume} {82}},\ \bibinfo {pages} {3045} (\bibinfo {year} {2010})}\BibitemShut {NoStop}%
\bibitem [{\citenamefont {Latzke}\ \emph {et~al.}(2015)\citenamefont {Latzke}, \citenamefont {Zhang}, \citenamefont {Suslu}, \citenamefont {Chang}, \citenamefont {Lin}, \citenamefont {Jeng}, \citenamefont {Tongay}, \citenamefont {Wu}, \citenamefont {Bansil},\ and\ \citenamefont {Lanzara}}]{latzke_electronic_2015}%
  \BibitemOpen
  \bibfield  {author} {\bibinfo {author} {\bibfnamefont {D.~W.}\ \bibnamefont {Latzke}}, \bibinfo {author} {\bibfnamefont {W.}~\bibnamefont {Zhang}}, \bibinfo {author} {\bibfnamefont {A.}~\bibnamefont {Suslu}}, \bibinfo {author} {\bibfnamefont {T.-R.}\ \bibnamefont {Chang}}, \bibinfo {author} {\bibfnamefont {H.}~\bibnamefont {Lin}}, \bibinfo {author} {\bibfnamefont {H.-T.}\ \bibnamefont {Jeng}}, \bibinfo {author} {\bibfnamefont {S.}~\bibnamefont {Tongay}}, \bibinfo {author} {\bibfnamefont {J.}~\bibnamefont {Wu}}, \bibinfo {author} {\bibfnamefont {A.}~\bibnamefont {Bansil}}, \ and\ \bibinfo {author} {\bibfnamefont {A.}~\bibnamefont {Lanzara}},\ }\href {\doibase 10.1103/PhysRevB.91.235202} {\bibfield  {journal} {\bibinfo  {journal} {Phys. Rev. B}\ }\textbf {\bibinfo {volume} {91}},\ \bibinfo {pages} {235202} (\bibinfo {year} {2015})}\BibitemShut {NoStop}%
\bibitem [{\citenamefont {Kim}\ \emph {et~al.}(2016)\citenamefont {Kim}, \citenamefont {Rhim}, \citenamefont {Kim}, \citenamefont {Kim},\ and\ \citenamefont {Park}}]{kim_determination_2016}%
  \BibitemOpen
  \bibfield  {author} {\bibinfo {author} {\bibfnamefont {B.~S.}\ \bibnamefont {Kim}}, \bibinfo {author} {\bibfnamefont {J.-W.}\ \bibnamefont {Rhim}}, \bibinfo {author} {\bibfnamefont {B.}~\bibnamefont {Kim}}, \bibinfo {author} {\bibfnamefont {C.}~\bibnamefont {Kim}}, \ and\ \bibinfo {author} {\bibfnamefont {S.~R.}\ \bibnamefont {Park}},\ }\href {\doibase 10.1038/srep36389} {\bibfield  {journal} {\bibinfo  {journal} {Scientific Reports}\ }\textbf {\bibinfo {volume} {6}},\ \bibinfo {pages} {36389} (\bibinfo {year} {2016})}\BibitemShut {NoStop}%
\bibitem [{\citenamefont {Alarab}\ \emph {et~al.}(2023)\citenamefont {Alarab}, \citenamefont {Minar}, \citenamefont {Constantinou}, \citenamefont {Nafday}, \citenamefont {Schmitt}, \citenamefont {Wang},\ and\ \citenamefont {Strocov}}]{alarab_k_2023}%
  \BibitemOpen
  \bibfield  {author} {\bibinfo {author} {\bibfnamefont {F.}~\bibnamefont {Alarab}}, \bibinfo {author} {\bibfnamefont {J.}~\bibnamefont {Minar}}, \bibinfo {author} {\bibfnamefont {P.}~\bibnamefont {Constantinou}}, \bibinfo {author} {\bibfnamefont {D.}~\bibnamefont {Nafday}}, \bibinfo {author} {\bibfnamefont {T.}~\bibnamefont {Schmitt}}, \bibinfo {author} {\bibfnamefont {X.}~\bibnamefont {Wang}}, \ and\ \bibinfo {author} {\bibfnamefont {V.~N.}\ \bibnamefont {Strocov}},\ }\href@noop {} {\bibfield  {journal} {\bibinfo  {journal} {arXiv:2310.11317}\ } (\bibinfo {year} {2023})}\BibitemShut {NoStop}%
\bibitem [{\citenamefont {Le~F\'evre}\ \emph {et~al.}(2024)\citenamefont {Le~F\'evre}, \citenamefont {Salazar}, \citenamefont {Jamet}, \citenamefont {Bertran}, \citenamefont {Bigi}, \citenamefont {Ourghi}, \citenamefont {Vergnaud}, \citenamefont {Pulkkinen}, \citenamefont {Minar}, \citenamefont {Jaouen},\ and\ \citenamefont {Rault}}]{lefevre_two_2024}%
  \BibitemOpen
  \bibfield  {author} {\bibinfo {author} {\bibfnamefont {P.}~\bibnamefont {Le~F\'evre}}, \bibinfo {author} {\bibfnamefont {R.}~\bibnamefont {Salazar}}, \bibinfo {author} {\bibfnamefont {M.}~\bibnamefont {Jamet}}, \bibinfo {author} {\bibfnamefont {F.}~\bibnamefont {Bertran}}, \bibinfo {author} {\bibfnamefont {C.}~\bibnamefont {Bigi}}, \bibinfo {author} {\bibfnamefont {A.}~\bibnamefont {Ourghi}}, \bibinfo {author} {\bibfnamefont {C.}~\bibnamefont {Vergnaud}}, \bibinfo {author} {\bibfnamefont {A.}~\bibnamefont {Pulkkinen}}, \bibinfo {author} {\bibfnamefont {J.}~\bibnamefont {Minar}}, \bibinfo {author} {\bibfnamefont {T.}~\bibnamefont {Jaouen}}, \ and\ \bibinfo {author} {\bibfnamefont {J.}~\bibnamefont {Rault}},\ }\href@noop {} {\bibfield  {journal} {\bibinfo  {journal} {arXiv:2407.03768}\ } (\bibinfo {year} {2024})}\BibitemShut {NoStop}%
\bibitem [{\citenamefont {Chang}\ \emph {et~al.}(2014)\citenamefont {Chang}, \citenamefont {Lin}, \citenamefont {Jeng},\ and\ \citenamefont {Bansil}}]{chang_thickness_2014}%
  \BibitemOpen
  \bibfield  {author} {\bibinfo {author} {\bibfnamefont {T.-R.}\ \bibnamefont {Chang}}, \bibinfo {author} {\bibfnamefont {H.}~\bibnamefont {Lin}}, \bibinfo {author} {\bibfnamefont {H.-T.}\ \bibnamefont {Jeng}}, \ and\ \bibinfo {author} {\bibfnamefont {A.}~\bibnamefont {Bansil}},\ }\href {\doibase 10.1038/srep06270} {\bibfield  {journal} {\bibinfo  {journal} {Scientific Reports}\ }\textbf {\bibinfo {volume} {4}},\ \bibinfo {pages} {6270} (\bibinfo {year} {2014})}\BibitemShut {NoStop}%
\bibitem [{\citenamefont {Mak}\ \emph {et~al.}(2010)\citenamefont {Mak}, \citenamefont {Lee}, \citenamefont {Hone}, \citenamefont {Shan},\ and\ \citenamefont {Heinz}}]{mak_atomically_2010}%
  \BibitemOpen
  \bibfield  {author} {\bibinfo {author} {\bibfnamefont {K.~F.}\ \bibnamefont {Mak}}, \bibinfo {author} {\bibfnamefont {C.}~\bibnamefont {Lee}}, \bibinfo {author} {\bibfnamefont {J.}~\bibnamefont {Hone}}, \bibinfo {author} {\bibfnamefont {J.}~\bibnamefont {Shan}}, \ and\ \bibinfo {author} {\bibfnamefont {T.~F.}\ \bibnamefont {Heinz}},\ }\href {\doibase 10.1103/PhysRevLett.105.136805} {\bibfield  {journal} {\bibinfo  {journal} {Phys. Rev. Lett.}\ }\textbf {\bibinfo {volume} {105}},\ \bibinfo {pages} {136805} (\bibinfo {year} {2010})}\BibitemShut {NoStop}%
\bibitem [{\citenamefont {Sun}\ \emph {et~al.}(2016)\citenamefont {Sun}, \citenamefont {Wang},\ and\ \citenamefont {Shuai}}]{sun_indirect_2016}%
  \BibitemOpen
  \bibfield  {author} {\bibinfo {author} {\bibfnamefont {Y.}~\bibnamefont {Sun}}, \bibinfo {author} {\bibfnamefont {D.}~\bibnamefont {Wang}}, \ and\ \bibinfo {author} {\bibfnamefont {Z.}~\bibnamefont {Shuai}},\ }\bibfield  {booktitle} {\emph {\bibinfo {booktitle} {The Journal of Physical Chemistry C}},\ }\href {\doibase 10.1021/acs.jpcc.6b08748} {\bibfield  {journal} {\bibinfo  {journal} {The Journal of Physical Chemistry C}\ }\textbf {\bibinfo {volume} {120}},\ \bibinfo {pages} {21866} (\bibinfo {year} {2016})}\BibitemShut {NoStop}%
\bibitem [{\citenamefont {Ernandes}\ \emph {et~al.}(2021)\citenamefont {Ernandes}, \citenamefont {Khalil}, \citenamefont {Almabrouk}, \citenamefont {Pierucci}, \citenamefont {Zheng}, \citenamefont {Avila}, \citenamefont {Dudin}, \citenamefont {Chaste}, \citenamefont {Oehler}, \citenamefont {Pala}, \citenamefont {Bisti}, \citenamefont {Brul{\'e}}, \citenamefont {Lhuillier}, \citenamefont {Pan},\ and\ \citenamefont {Ouerghi}}]{ernandes_indirect_2021}%
  \BibitemOpen
  \bibfield  {author} {\bibinfo {author} {\bibfnamefont {C.}~\bibnamefont {Ernandes}}, \bibinfo {author} {\bibfnamefont {L.}~\bibnamefont {Khalil}}, \bibinfo {author} {\bibfnamefont {H.}~\bibnamefont {Almabrouk}}, \bibinfo {author} {\bibfnamefont {D.}~\bibnamefont {Pierucci}}, \bibinfo {author} {\bibfnamefont {B.}~\bibnamefont {Zheng}}, \bibinfo {author} {\bibfnamefont {J.}~\bibnamefont {Avila}}, \bibinfo {author} {\bibfnamefont {P.}~\bibnamefont {Dudin}}, \bibinfo {author} {\bibfnamefont {J.}~\bibnamefont {Chaste}}, \bibinfo {author} {\bibfnamefont {F.}~\bibnamefont {Oehler}}, \bibinfo {author} {\bibfnamefont {M.}~\bibnamefont {Pala}}, \bibinfo {author} {\bibfnamefont {F.}~\bibnamefont {Bisti}}, \bibinfo {author} {\bibfnamefont {T.}~\bibnamefont {Brul{\'e}}}, \bibinfo {author} {\bibfnamefont {E.}~\bibnamefont {Lhuillier}}, \bibinfo {author} {\bibfnamefont {A.}~\bibnamefont {Pan}}, \ and\ \bibinfo {author} {\bibfnamefont {A.}~\bibnamefont {Ouerghi}},\ }\href {\doibase 10.1038/s41699-020-00187-9} {\bibfield
  {journal} {\bibinfo  {journal} {npj 2D Materials and Applications}\ }\textbf {\bibinfo {volume} {5}},\ \bibinfo {pages} {7} (\bibinfo {year} {2021})}\BibitemShut {NoStop}%
\bibitem [{\citenamefont {Luttinger}\ and\ \citenamefont {Ward}(1960)}]{luttinger_1960}%
  \BibitemOpen
  \bibfield  {author} {\bibinfo {author} {\bibfnamefont {J.~M.}\ \bibnamefont {Luttinger}}\ and\ \bibinfo {author} {\bibfnamefont {J.~C.}\ \bibnamefont {Ward}},\ }\href {\doibase 10.1103/PhysRev.118.1417} {\bibfield  {journal} {\bibinfo  {journal} {Phys. Rev.}\ }\textbf {\bibinfo {volume} {118}},\ \bibinfo {pages} {1417} (\bibinfo {year} {1960})}\BibitemShut {NoStop}%
\bibitem [{\citenamefont {Clark}\ \emph {et~al.}()\citenamefont {Clark}, \citenamefont {Azhar}, \citenamefont {Chambers}, \citenamefont {McEwen}, \citenamefont {Vu}, \citenamefont {Bhuiyan}, \citenamefont {Belosludov}, \citenamefont {A.}, \citenamefont {Jozwiak}, \citenamefont {Rotenberg}, \citenamefont {Lee}, \citenamefont {Mao}, \citenamefont {Balakrishnan}, \citenamefont {Mazzola}, \citenamefont {Harmer}, \citenamefont {Fuhrer}, \citenamefont {Bahramy},\ and\ \citenamefont {Edmonds}}]{clark_dimensional_2025}%
  \BibitemOpen
  \bibfield  {author} {\bibinfo {author} {\bibfnamefont {O.~J.}\ \bibnamefont {Clark}}, \bibinfo {author} {\bibfnamefont {A.}~\bibnamefont {Azhar}}, \bibinfo {author} {\bibfnamefont {B.~A.}\ \bibnamefont {Chambers}}, \bibinfo {author} {\bibfnamefont {D.}~\bibnamefont {McEwen}}, \bibinfo {author} {\bibfnamefont {T.-H.-Y.}\ \bibnamefont {Vu}}, \bibinfo {author} {\bibfnamefont {M.~T.~H.}\ \bibnamefont {Bhuiyan}}, \bibinfo {author} {\bibfnamefont {R.~H.}\ \bibnamefont {Belosludov}}, \bibinfo {author} {\bibfnamefont {B.}~\bibnamefont {A.}}, \bibinfo {author} {\bibfnamefont {C.}~\bibnamefont {Jozwiak}}, \bibinfo {author} {\bibfnamefont {E.}~\bibnamefont {Rotenberg}}, \bibinfo {author} {\bibfnamefont {S.~H.}\ \bibnamefont {Lee}}, \bibinfo {author} {\bibfnamefont {Z.}~\bibnamefont {Mao}}, \bibinfo {author} {\bibfnamefont {G.}~\bibnamefont {Balakrishnan}}, \bibinfo {author} {\bibfnamefont {F.}~\bibnamefont {Mazzola}}, \bibinfo {author} {\bibfnamefont {S.~L.}\ \bibnamefont {Harmer}}, \bibinfo {author} {\bibfnamefont
  {M.~S.}\ \bibnamefont {Fuhrer}}, \bibinfo {author} {\bibfnamefont {M.~S.}\ \bibnamefont {Bahramy}}, \ and\ \bibinfo {author} {\bibfnamefont {M.~T.}\ \bibnamefont {Edmonds}},\ }\href@noop {} {\bibinfo  {journal} {arXiv:2503.17947}\ }\BibitemShut {NoStop}%
\bibitem [{\citenamefont {Xi}\ \emph {et~al.}(2015)\citenamefont {Xi}, \citenamefont {Zhao}, \citenamefont {Wang}, \citenamefont {Berger}, \citenamefont {Forr{\'o}}, \citenamefont {Shan},\ and\ \citenamefont {Mak}}]{xi_strongly_2015}%
  \BibitemOpen
\bibfield  {journal} {  }\bibfield  {author} {\bibinfo {author} {\bibfnamefont {X.}~\bibnamefont {Xi}}, \bibinfo {author} {\bibfnamefont {L.}~\bibnamefont {Zhao}}, \bibinfo {author} {\bibfnamefont {Z.}~\bibnamefont {Wang}}, \bibinfo {author} {\bibfnamefont {H.}~\bibnamefont {Berger}}, \bibinfo {author} {\bibfnamefont {L.}~\bibnamefont {Forr{\'o}}}, \bibinfo {author} {\bibfnamefont {J.}~\bibnamefont {Shan}}, \ and\ \bibinfo {author} {\bibfnamefont {K.~F.}\ \bibnamefont {Mak}},\ }\href {\doibase 10.1038/nnano.2015.143} {\bibfield  {journal} {\bibinfo  {journal} {Nature Nanotechnology}\ }\textbf {\bibinfo {volume} {10}},\ \bibinfo {pages} {765} (\bibinfo {year} {2015})}\BibitemShut {NoStop}%
\bibitem [{\citenamefont {Dvir}\ \emph {et~al.}(2018)\citenamefont {Dvir}, \citenamefont {Massee}, \citenamefont {Attias}, \citenamefont {Khodas}, \citenamefont {Aprili}, \citenamefont {Quay},\ and\ \citenamefont {Steinberg}}]{dvir_spectroscopy_2018}%
  \BibitemOpen
  \bibfield  {author} {\bibinfo {author} {\bibfnamefont {T.}~\bibnamefont {Dvir}}, \bibinfo {author} {\bibfnamefont {F.}~\bibnamefont {Massee}}, \bibinfo {author} {\bibfnamefont {L.}~\bibnamefont {Attias}}, \bibinfo {author} {\bibfnamefont {M.}~\bibnamefont {Khodas}}, \bibinfo {author} {\bibfnamefont {M.}~\bibnamefont {Aprili}}, \bibinfo {author} {\bibfnamefont {C.~H.~L.}\ \bibnamefont {Quay}}, \ and\ \bibinfo {author} {\bibfnamefont {H.}~\bibnamefont {Steinberg}},\ }\href {\doibase 10.1038/s41467-018-03000-w} {\bibfield  {journal} {\bibinfo  {journal} {Nature Communications}\ }\textbf {\bibinfo {volume} {9}},\ \bibinfo {pages} {598} (\bibinfo {year} {2018})}\BibitemShut {NoStop}%
\bibitem [{\citenamefont {Hamill}\ \emph {et~al.}(2021)\citenamefont {Hamill}, \citenamefont {Heischmidt}, \citenamefont {Sohn}, \citenamefont {Shaffer}, \citenamefont {Tsai}, \citenamefont {Zhang}, \citenamefont {Xi}, \citenamefont {Suslov}, \citenamefont {Berger}, \citenamefont {Forr{\'o}}, \citenamefont {Burnell}, \citenamefont {Shan}, \citenamefont {Mak}, \citenamefont {Fernandes}, \citenamefont {Wang},\ and\ \citenamefont {Pribiag}}]{hamill_two_2021}%
  \BibitemOpen
  \bibfield  {author} {\bibinfo {author} {\bibfnamefont {A.}~\bibnamefont {Hamill}}, \bibinfo {author} {\bibfnamefont {B.}~\bibnamefont {Heischmidt}}, \bibinfo {author} {\bibfnamefont {E.}~\bibnamefont {Sohn}}, \bibinfo {author} {\bibfnamefont {D.}~\bibnamefont {Shaffer}}, \bibinfo {author} {\bibfnamefont {K.-T.}\ \bibnamefont {Tsai}}, \bibinfo {author} {\bibfnamefont {X.}~\bibnamefont {Zhang}}, \bibinfo {author} {\bibfnamefont {X.}~\bibnamefont {Xi}}, \bibinfo {author} {\bibfnamefont {A.}~\bibnamefont {Suslov}}, \bibinfo {author} {\bibfnamefont {H.}~\bibnamefont {Berger}}, \bibinfo {author} {\bibfnamefont {L.}~\bibnamefont {Forr{\'o}}}, \bibinfo {author} {\bibfnamefont {F.~J.}\ \bibnamefont {Burnell}}, \bibinfo {author} {\bibfnamefont {J.}~\bibnamefont {Shan}}, \bibinfo {author} {\bibfnamefont {K.~F.}\ \bibnamefont {Mak}}, \bibinfo {author} {\bibfnamefont {R.~M.}\ \bibnamefont {Fernandes}}, \bibinfo {author} {\bibfnamefont {K.}~\bibnamefont {Wang}}, \ and\ \bibinfo {author} {\bibfnamefont {V.~S.}\
  \bibnamefont {Pribiag}},\ }\href {\doibase 10.1038/s41567-021-01219-x} {\bibfield  {journal} {\bibinfo  {journal} {Nature Physics}\ }\textbf {\bibinfo {volume} {17}},\ \bibinfo {pages} {949} (\bibinfo {year} {2021})}\BibitemShut {NoStop}%
\bibitem [{\citenamefont {Xi}\ \emph {et~al.}(2016)\citenamefont {Xi}, \citenamefont {Wang}, \citenamefont {Zhao}, \citenamefont {Park}, \citenamefont {Law}, \citenamefont {Berger}, \citenamefont {Forr{\'o}}, \citenamefont {Shan},\ and\ \citenamefont {Mak}}]{xi_ising_2016}%
  \BibitemOpen
  \bibfield  {author} {\bibinfo {author} {\bibfnamefont {X.}~\bibnamefont {Xi}}, \bibinfo {author} {\bibfnamefont {Z.}~\bibnamefont {Wang}}, \bibinfo {author} {\bibfnamefont {W.}~\bibnamefont {Zhao}}, \bibinfo {author} {\bibfnamefont {J.-H.}\ \bibnamefont {Park}}, \bibinfo {author} {\bibfnamefont {K.~T.}\ \bibnamefont {Law}}, \bibinfo {author} {\bibfnamefont {H.}~\bibnamefont {Berger}}, \bibinfo {author} {\bibfnamefont {L.}~\bibnamefont {Forr{\'o}}}, \bibinfo {author} {\bibfnamefont {J.}~\bibnamefont {Shan}}, \ and\ \bibinfo {author} {\bibfnamefont {K.~F.}\ \bibnamefont {Mak}},\ }\href {\doibase 10.1038/nphys3538} {\bibfield  {journal} {\bibinfo  {journal} {Nature Physics}\ }\textbf {\bibinfo {volume} {12}},\ \bibinfo {pages} {139} (\bibinfo {year} {2016})}\BibitemShut {NoStop}%
\bibitem [{\citenamefont {Yoshizawa}\ \emph {et~al.}(2024)\citenamefont {Yoshizawa}, \citenamefont {Sagisaka},\ and\ \citenamefont {Sakata}}]{yoshizawa_visualization_2024}%
  \BibitemOpen
  \bibfield  {author} {\bibinfo {author} {\bibfnamefont {S.}~\bibnamefont {Yoshizawa}}, \bibinfo {author} {\bibfnamefont {K.}~\bibnamefont {Sagisaka}}, \ and\ \bibinfo {author} {\bibfnamefont {H.}~\bibnamefont {Sakata}},\ }\href {\doibase 10.1103/PhysRevLett.132.056401} {\bibfield  {journal} {\bibinfo  {journal} {Phys. Rev. Lett.}\ }\textbf {\bibinfo {volume} {132}},\ \bibinfo {pages} {056401} (\bibinfo {year} {2024})}\BibitemShut {NoStop}%
\bibitem [{\citenamefont {Sato}\ and\ \citenamefont {Ando}(2017)}]{sato_topological_2017}%
  \BibitemOpen
  \bibfield  {author} {\bibinfo {author} {\bibfnamefont {M.}~\bibnamefont {Sato}}\ and\ \bibinfo {author} {\bibfnamefont {Y.}~\bibnamefont {Ando}},\ }\href@noop {} {\bibfield  {journal} {\bibinfo  {journal} {Reports on Progress in Physics}\ }\textbf {\bibinfo {volume} {80}},\ \bibinfo {pages} {076501} (\bibinfo {year} {2017})}\BibitemShut {NoStop}%
\bibitem [{\citenamefont {Shao}\ \emph {et~al.}(2016)\citenamefont {Shao}, \citenamefont {Yu}, \citenamefont {Lan}, \citenamefont {Shi}, \citenamefont {Li}, \citenamefont {Zheng}, \citenamefont {Zhu}, \citenamefont {Li}, \citenamefont {Amiri},\ and\ \citenamefont {Wang}}]{shao_strong_2016}%
  \BibitemOpen
  \bibfield  {author} {\bibinfo {author} {\bibfnamefont {Q.}~\bibnamefont {Shao}}, \bibinfo {author} {\bibfnamefont {G.}~\bibnamefont {Yu}}, \bibinfo {author} {\bibfnamefont {Y.-W.}\ \bibnamefont {Lan}}, \bibinfo {author} {\bibfnamefont {Y.}~\bibnamefont {Shi}}, \bibinfo {author} {\bibfnamefont {M.-Y.}\ \bibnamefont {Li}}, \bibinfo {author} {\bibfnamefont {C.}~\bibnamefont {Zheng}}, \bibinfo {author} {\bibfnamefont {X.}~\bibnamefont {Zhu}}, \bibinfo {author} {\bibfnamefont {L.-J.}\ \bibnamefont {Li}}, \bibinfo {author} {\bibfnamefont {P.~K.}\ \bibnamefont {Amiri}}, \ and\ \bibinfo {author} {\bibfnamefont {K.~L.}\ \bibnamefont {Wang}},\ }\href {\doibase 10.1021/acs.nanolett.6b03300} {\bibfield  {journal} {\bibinfo  {journal} {Nano Letters}\ }\textbf {\bibinfo {volume} {16}},\ \bibinfo {pages} {7514} (\bibinfo {year} {2016})}\BibitemShut {NoStop}%
\bibitem [{\citenamefont {He}\ \emph {et~al.}(2025)\citenamefont {He}, \citenamefont {Yin}, \citenamefont {Zhang}, \citenamefont {Chen}, \citenamefont {Peng}, \citenamefont {Shan}, \citenamefont {Zhao},\ and\ \citenamefont {Gao}}]{he_all_2025}%
  \BibitemOpen
  \bibfield  {author} {\bibinfo {author} {\bibfnamefont {S.}~\bibnamefont {He}}, \bibinfo {author} {\bibfnamefont {C.}~\bibnamefont {Yin}}, \bibinfo {author} {\bibfnamefont {L.}~\bibnamefont {Zhang}}, \bibinfo {author} {\bibfnamefont {Y.}~\bibnamefont {Chen}}, \bibinfo {author} {\bibfnamefont {H.}~\bibnamefont {Peng}}, \bibinfo {author} {\bibfnamefont {A.}~\bibnamefont {Shan}}, \bibinfo {author} {\bibfnamefont {L.}~\bibnamefont {Zhao}}, \ and\ \bibinfo {author} {\bibfnamefont {L.}~\bibnamefont {Gao}},\ }\href {\doibase https://doi.org/10.1016/j.jmst.2024.08.055} {\bibfield  {journal} {\bibinfo  {journal} {Journal of Materials Science \& Technology}\ }\textbf {\bibinfo {volume} {219}},\ \bibinfo {pages} {205} (\bibinfo {year} {2025})}\BibitemShut {NoStop}%
\bibitem [{\citenamefont {Vu}\ \emph {et~al.}(2024)\citenamefont {Vu}, \citenamefont {Nguyen}, \citenamefont {Kim}, \citenamefont {Do}, \citenamefont {Dat},\ and\ \citenamefont {Yu}}]{vu_synthesis_2024}%
  \BibitemOpen
  \bibfield  {author} {\bibinfo {author} {\bibfnamefont {V.~T.}\ \bibnamefont {Vu}}, \bibinfo {author} {\bibfnamefont {M.~C.}\ \bibnamefont {Nguyen}}, \bibinfo {author} {\bibfnamefont {W.~K.}\ \bibnamefont {Kim}}, \bibinfo {author} {\bibfnamefont {V.~D.}\ \bibnamefont {Do}}, \bibinfo {author} {\bibfnamefont {V.~K.}\ \bibnamefont {Dat}}, \ and\ \bibinfo {author} {\bibfnamefont {W.~J.}\ \bibnamefont {Yu}},\ }\href {\doibase https://doi.org/10.1002/sstr.202300401} {\bibfield  {journal} {\bibinfo  {journal} {Small Structures}\ }\textbf {\bibinfo {volume} {5}},\ \bibinfo {pages} {2300401} (\bibinfo {year} {2024})}\BibitemShut {NoStop}%
\bibitem [{\citenamefont {Fang}\ \emph {et~al.}(2023)\citenamefont {Fang}, \citenamefont {Lin}, \citenamefont {Zhang}, \citenamefont {Thompson}, \citenamefont {Xiao}, \citenamefont {Sun}, \citenamefont {Malic}, \citenamefont {Dash},\ and\ \citenamefont {Wieczorek}}]{fang_localization_2023}%
  \BibitemOpen
  \bibfield  {author} {\bibinfo {author} {\bibfnamefont {H.}~\bibnamefont {Fang}}, \bibinfo {author} {\bibfnamefont {Q.}~\bibnamefont {Lin}}, \bibinfo {author} {\bibfnamefont {Y.}~\bibnamefont {Zhang}}, \bibinfo {author} {\bibfnamefont {J.}~\bibnamefont {Thompson}}, \bibinfo {author} {\bibfnamefont {S.}~\bibnamefont {Xiao}}, \bibinfo {author} {\bibfnamefont {Z.}~\bibnamefont {Sun}}, \bibinfo {author} {\bibfnamefont {E.}~\bibnamefont {Malic}}, \bibinfo {author} {\bibfnamefont {S.~P.}\ \bibnamefont {Dash}}, \ and\ \bibinfo {author} {\bibfnamefont {W.}~\bibnamefont {Wieczorek}},\ }\href {\doibase 10.1038/s41467-023-42710-8} {\bibfield  {journal} {\bibinfo  {journal} {Nature Communications}\ }\textbf {\bibinfo {volume} {14}},\ \bibinfo {pages} {6910} (\bibinfo {year} {2023})}\BibitemShut {NoStop}%
\bibitem [{Fli()}]{FlindersNanoESCA}%
  \BibitemOpen
  \href {\doibase 10.25957/flinders.nanoesca3} {\ 10.25957/flinders.nanoesca3}\BibitemShut {NoStop}%
\end{thebibliography}%

\noindent {\bf Acknowledgments}

\noindent{
O. J. C. thanks M. Saeed Bahramy for useful discussions.  
O. J. C., M. S. F. and M. T. E. knowledge funding support from ARC Discovery Project DP200101345. M. T. E. acknowledges funding support from ARC Future Fellowship FT2201000290. O. J. C. acknowledges travel funding provided by the International Synchrotron Access Program (ISAP) managed by the Australian Synchrotron, part of ANSTO, and funded by the Australian Government. This research used resources of the Advanced Light Source, which is a DOE Office of Science User Facility under contract no. DE-AC02-05CH11231. This work was performed in part at the Melbourne Centre for Nanofabrication (MCN) in the Victorian Node of the Australian National Fabrication Facility (ANFF). The authors acknowledge the facilities, and the scientific and technical assistance of Microscopy Australia (ROR: 042mm0k03) enabled by NCRIS and the government of South Australia at Flinders Microscopy and Microanalysis (ROR: 04z91ja70), Flinders University (ROR: 01kpzv902).

} 

\

\noindent {\bf Author Contributions}
\noindent {O.J.C. constructed the heterostructures.  G. B. grew the NbSe$_2$ single crystals. O. J. C., T.-H.-Y. V., F. M., and S. S. performed photoemission measurements, supported by A. B., C. J. and E. R.. O. J. C. and B. A. C. performed the nano-ESCA measurements. O. J. C. performed the photoluminsenscence measurements. B. A. C. and S. L. H. proved access to the nano-ESCA instrument at Flinders University, South Australia.
O. J. C. wrote the manuscript with significant contributions from M. S. F. and M. T. E. and further contributions from all other authors. O. J. C. was responsible for project planning and direction.  }

\

\noindent {\bf Data Availability}

\noindent {Data are available from the authors upon reasonable request.}

\phantom{xxxx}

\end{document}